\documentclass[twocolumn]{article}
\usepackage{amssymb}
\usepackage{mathptmx}
\usepackage{times}
\usepackage{amsthm,amsmath,amssymb}
\usepackage{subfig}
\usepackage{algorithm}
\usepackage{algpseudocode}
\usepackage{enumerate}
\usepackage[all]{xy}
\usepackage{tikz-cd}
\usepackage{url}
\usepackage{amsfonts}
\usepackage{mathrsfs}
\usepackage{comment}
\usepackage[switch]{lineno} 
\usepackage{graphicx}
\usepackage[margin=.6 in]{geometry}

\newtheorem{theorem}{Theorem}[section]
\newtheorem{lemma}[theorem]{Lemma}

\newcommand{\bx}{\mathbf{x}}
\newcommand{\bc}{\mathbf{c}}
\newcommand{\bs}{\mathbf{s}}

\newcommand{\I}{\mathbb{I}}
\newcommand{\X}{\mathbb{X}}
\newcommand{\J}{\mathbb{J}}

\newcommand{\R}{\mathbb{R}}

\newcommand{\Sp}{\mathbb{S}}
\newcommand{\Int}[1]{{#1}^{\circ}}
\newcommand{\Bd}[1]{{#1}^{\partial}}

\newcommand{\RS}{\mathbb{W}_f}
\newcommand{\JS}{\mathfrak{J}_f}
\newcommand{\Ji}{\Int{\mathbb{J}}_f}
\newcommand{\Jb}{\Bd{\mathbb{J}}_f}
\newcommand{\Jf}{\mathbb{J}_f}
\newcommand{\JCN}[2]{JCN#1\,#2}
\newcommand{\MDRG}{\mathbb{M}_f}
\newcommand{\RG}{\mathbb{R}_{f_i}}

\newcommand{\RK}{\mathbb{K}_f}

\newcommand{\figref}[1]{Figure \ref{fig:#1}}
\newcommand{\twofigref}[2]{Figures \ref{fig:#1} and \ref{fig:#2}}
\newcommand{\threefigref}[3]{Figures \ref{fig:#1}, \ref{fig:#2} and \ref{fig:#3}}
\newcommand{\secref}[1]{Section~\ref{sec:#1}}

\newcommand{\subsecref}[1]{Subsection \ref{sec:#1}}

\newcommand{\tabref}[1]{Table \ref{tab:#1}}
\newcommand{\lemref}[1]{Lemma \ref{lem:#1}}
\newcommand{\algoref}[1]{Algorithm \ref{alg:#1}}
\newcommand{\etal}{et~al.}

\newtheorem{dfn}{Definition}[section]

{\begin{Sbox}\begin{minipage}}%
{\end{minipage}\end{Sbox}\fbox{\TheSbox}}
\renewenvironment{proof}{{\bf Proof. }}{\hspace*{\fill}$\Box$\bigskip\noindent}

\begin{document}
\title{\textbf{ Multivariate Topology Simplification}}


\author{Amit Chattopadhyay \thanks{ School of Computing, University of Leeds, 
  Leeds, UK. {\tt A.Chattopadhyay@leeds.ac.uk}}
\and
Hamish Carr \thanks{ School of Computing, University of Leeds, 
  Leeds, UK. {\tt H.Carr@leeds.ac.uk}}
\and
David Duke\thanks{ School of Computing, University of Leeds, 
  Leeds, UK. {\tt D.J.Duke@leeds.ac.uk}}
\and
Zhao Geng \thanks{ School of Computing, University of Leeds, 
  Leeds, UK. {\tt Z.Geng@leeds.ac.uk}}
\and
Osamu Saeki \thanks{Institute of Mathematics for Industry, Kyushu
  University, Japan. {\tt saeki@imi.kyushu-u.ac.jp}}
}

\maketitle

\begin{abstract}
Topological simplification of scalar and vector fields is well-established as 
an effective method for analysing and visualising complex data sets. For
multi-field data, topological analysis requires simultaneous advances
both mathematically and computationally. We propose a robust
multivariate topology simplification method based on ``lip''-pruning from the Reeb Space.
Mathematically, we show that the projection of the Jacobi Set of multivariate data into the Reeb Space produces
a Jacobi Structure that separates the Reeb Space into simple components.  We also show that the dual
graph of these components gives rise to a Reeb Skeleton that has properties similar
to the scalar contour tree and Reeb Graph, for topologically simple
domains.  We then introduce a range measure to give a scaling-invariant total
ordering of the components or features that can be used for simplification. Computationally, we show
how to compute Jacobi Structure, Reeb Skeleton, Range and Geometric Measures
in the Joint Contour Net (an approximation of the Reeb Space) and that these can be used for
visualisation similar to the contour tree or Reeb Graph.
\end{abstract}

\paragraph*{keyword}
Simplification, Topology, Multi-Field, Reeb Space, Joint Contour Net,
Multi-Dimensional Reeb Graph, Reeb Skeleton

\section{Introduction} 
\label{sec:intro}
Scientific data is often complex in nature and difficult to visualise.  As a result, 
analytic tools have become increasingly prominent in scientific visualisation,
and in particular topological analysis.  While earlier work dealt primarily with
scalar data~\cite{2003-Chiang-Simplification, 2004-Carr-simplification, 2012-Tierny-tvcg}, 
multivariate topological analysis in the form of the Reeb
Space~\cite{2004-Saeki, 2008-edels-reebspace}
has started to become feasible using a quantised approximation called the Joint 
Contour Net (JCN)~\cite{2013-Carr-TVCG}.

Prior experience in scalar and vector topology shows that simplification of
topological structures is required, as real data sets are often noisy and 
complex.  Although most of the work required is practical and algorithmic in 
nature, mathematical formalisms are also needed, in this case based on fiber 
analysis, in the same way that Reeb Graphs and contour trees rely on Morse theory.
This paper therefore:
\begin{enumerate}\itemsep=.8pt
\item	Clarifies relationships between the Reeb Space of a multivariate map $f$, 
the Jacobi Set of $f$, and fiber topology,
\item	Introduces the \emph{Jacobi Structure} in the Reeb Space that decomposes the
Reeb Space into \emph{regular} and \emph{singular} components equivalent to edges and vertices
in the Reeb Graph, then reduces it further to a \textit{Reeb Skeleton}, 
\item	Proves that Reeb Spaces for topologically simple domains have simple structures
with properties analogous to properties of the contour tree, allowing
\emph{lip-pruning} based simplification,
\item	Introduces the \emph{range measure} and other geometric measures
  for a total ordering of regular components of the Reeb Space,
\item	Describes an algorithm that extracts the Jacobi Structure from the Joint Contour Net 
using a \textit{Multi-dimensional Reeb Graph} (MDRG) and computes the Reeb Skeleton, 
and
\item	Simplifies the Reeb Skeleton and the corresponding Reeb Space
  computing the range and other geometric measures using the Joint Contour Net. 
\end{enumerate}

To clarify the relationships between the newly introduced
data-structures in the current paper, note
that the JCN is an approximation of the Reeb Space. We compute 
a MDRG from the JCN. The critical nodes of
the MDRG form the Jacobi Structure of the JCN. The  Jacobi Structure then
separates the JCN into regular and singular components. The dual graph
of such components gives a Reeb Skeleton which is used in the
multivariate topology simplification.

As a result, much of this paper addresses the theoretical machinery for simplification of the 
Reeb Space and its approximation, the Joint Contour Net. ~\secref{PreviousWork} reviews 
relevant background material on simplification, followed by a more
detailed review of the fiber topology, Jacobi Set and Reeb Space in~\secref{Background}.  \secref{Theory} provides
theoretical analysis and results needed for the lip-simplification of
the Reeb Space. For simple domains, the Reeb Space can have detachable
(lip) components: this is used in \secref{Simplification} to generalise leaf-pruning simplification from the
contour tree to the Reeb Space.  Once this has been done, we introduce a range persistence and
other geometric measures to govern the simplification process.  

In \secref{SimplifyingJCN}, we give an algorithm for simplifying the Joint Contour 
Net (an approximation of the Reeb Space). We start by building a hierarchical structure called the Multi-Dimensional Reeb
Graph (MDRG) that captures the Jacobi Structure of the Joint Contour Net,
and then show how to reduce the JCN to a Reeb Skeleton - a graph with properties similar to a contour tree. 
In \secref{Implementation}, we illustrate these reductions first with 
analytic data where the correct solution is known \emph{a priori},
then for a real data from the nuclear physics. As part of this, we provide performance figures and other implementation details in 
\secref{Implementation}, then draw conclusions and lay out a road map
for further work in \secref{Conclusions}.


\section{Previous Work}
\label{sec:PreviousWork}
Topology-based simplification aims to reduce the topological
complexity of the underlying data. There are different ways to
measure such topological complexity depending on the nature of the
underlying data. Here we mention some well-known approaches from the
literature for measuring the topological complexity and their
simplification procedure. 

\subsection*{Scalar Field Simplification.} 
The topological complexity of the scalar field data is measured in
terms of the number of critical points and their connectivities -
captured by its Reeb Graph or contour tree.  Another way to capture
the topological complexity of the scalar field is by computing the Morse-Smale complex of the corresponding
gradient field. Therefore, the topological simplification
in this case is driven by reducing the number of critical points via
simplification of the Reeb Graph/ contour tree or the Morse-Smale complex.
Carr et al. \cite{2004-Carr-simplification} describe a method for associating local geometric measures such
  as the surface area and the contained volume of contours with the contour
  tree and then simplifying the contour tree by suppressing the minor
  \emph{topological features} of the data. Note that  a feature is any
  prominent or distinctive part or quality that characterises the data
  and topological features captures the topological phenomena of the underlying data.  Wood et al. \cite{2004-Wood-excessTopo} give a
  Reeb Graph based simplification strategy for removing the excess
  topology created by unwanted handles in an isosurface using a
  measure for computing the handle-size in the isosurface and associating them
  with the loops of the Reeb Graph. Gyulassy et al. \cite{2006-Gyulassy} describe a
  technique for simplifying a three-dimensional scalar field by
  repeatedly removing pair of critical points from the Morse-Smale
  complex of its gradient field, by repeated application of a critical-point
  simplification operation. Mathematically, the simplification of
  ``lips'' proposed in this paper is a direct generalization of this
  idea (for scalar fields) to multi-fields. 
Luo et al \cite{2009-Luo-Jacobi} describe a method for computing and simplifying gradients and critical points of a function from a
point cloud. Tierny et al. \cite{2012-Tierny-tvcg} present a
  combinatorial algorithm for simplifying the topology of a scalar field on a surface
  by approximating with a simpler scalar field having a subset of
  critical points of the given field, while guaranteeing a small error distance
  between the fields.

The topological complexity of a point cloud
data can be measured by its homology. For a point cloud data  in
$\mathbb{R}^3$ this is expressed by the topological invariants, such as the Betti
numbers corresponding to a simplicial complex of the point cloud
 - denoted by $\beta_0$ (number of connected components), $\beta_1$ (number of
 tunnels or 1-dimensional holes) and
$\beta_2$ (number of voids or 2-dimensional holes).  The $i$-th Betti number represents the
rank of the $i$-th homology group ($i=0, 1, 2$). Edelsbrunner et al. \cite{2002-Edels-Persist} introduce the
idea of persistence homology for the topological simplification of a
point cloud by reducing the Betti numbers using a filtration
technique. Cohen-Steiner et al. \cite{2007-Cohen-Steiner} extend the persistence diagram
for scalar functions on topological spaces and analyze its
stability. 

\subsection*{Mesh Simplification.} Mesh-simplification is well-known 
in the computational geometry and graphics community. Topological complexity of a mesh can be
determined by its genus. Guskov et al. \cite{2001-Guskov-topologicalnoise} remove
the unnecessary topological noise from 
meshes of laser scanner data by reducing their genera. 
Nooruddin et al. \cite{2003-Nooruddin-tvcg} give a voxel-based simplification and repair method of polygonal models
using a volumetric morphological operation. Ni et al. \cite{2004-Ni-ATOG}
generate a fair Morse function for extracting the topological
structure of a surface mesh by user-controlled number and
configuration of critical points. Hoppe et al. \cite{1993-Hoppe-mesh-opti} describe
a energy-minimization technique for generating an optimal mesh by
reducing the number of vertices from a given mesh. Also Hoppe et al. \cite{1996-Hoppe-progressive-mesh, 1997-Hoppe-progressive-mesh} give a new progressive mesh
representation, a new scheme for storing and transmitting arbitrary
triangle meshes, and their simplification technique. Chiang et al. \cite{2003-Chiang-Simplification}
describe a technique of progressive simplification of tetrahedral
meshes preserving isosurface topologies. Their method works in two
stages - first they segment the volume data into
topological-equivalence regions and in the second step they simplify
each topological-equivalence region independently by edge collapsing, preserving the
iso-surface topologies. There are many cost-driven methods of mesh-simplification
(in the literature) which attempt to measure only the cost of each individual edge collapse
and the entire simplification process is considered as a sequence of
steps of increasing cost
\cite{1998-Dey-topologypreserving, 1998-LindstromT,
  1999-Lindstrom-Memless, 1997-GarlandH}.

\subsection*{Vector Field Simplification.} Topology based
  methods for vector field simplification are based on the
  idea of \textit{singularity pair cancellation} to reduce the number
  of singularities and thus the topological complexity. This method
  iteratively eliminates suitable pairs of singularities with opposite
  Poincar\'e-Hopf indices so that total sum of the indices remain
  invariant to keep the global structure of the field the same.
  This idea has been exploited in 
  \cite{2001-Tricoche-vec-simpli, 2006-Zhang-vec, 2011-Reininghaus-Visweek}.
There are also non-topology based methods for vector-field
  simplification which are mainly based on
  smoothing operations. Smoothing operations reduce vector and tensor-field
  complexity and remove large percentage of
  singularities. Polthier et al. \cite{2003-Polthier} apply Laplacian 
 smoothing on the potential of a
  vector-field. Tong et al. \cite{2003-Tong} decompose a vector field into three
  components: curl free, divergence free and harmonic. Each component
  is smoothed individually and results are summed to obtain simplified
  vector field.

\subsection*{Multi-Field Simplification.} 
To the best of our knowledge, until now there is no prior work
 on topology-based simplification of general multi-field data. 
All those techniques, cited so far, for simplifying scalar fields,
meshes and vector fields are not directly applicable in case of
multi-fields, mainly because the computation of the equivalent tools
such as, Jacobi Set \cite{2004-edels-localglobal}, Reeb Space \cite{2008-edels-reebspace} are
not well-developed.
A generalization of the persistence homology is proven to be difficult
for the multi-fields \cite{2007-carlsson-persist-multi}. 
However, few attempts have been made for
simplifying the Jacobi Sets in restrictive cases. 
Snyder et al. \cite{2004-snyder} give two metrics for
measuring persistence of the  Jacobi Sets.  Bremer et al. \cite{2007-Bremer} describe a
method for noise removal from the Jacobi Sets of time varying data.
Suthambhara et al. \cite{2009-Nataraj-Jacobi} give a technique for the
Jacobi Set simplification of bivariate fields based on simplification
of the Reeb Graphs of their comparison measures. 
Huettenberger et al. propose multi-field
simplification method using Pareto sets \cite{2013-Huettenberger-pareto,
  2014-Huettenberger-tvcg}. However, these methods lack 
mathematical justification for simplifying the corresponding input multi-fields and
work mostly for bivariate data.
In a similar context, Bhatia et al. \cite{2013-Bhatia} provide a simplification
method by generalising the critical point cancellation of scalar
functions to the  Jacobi Sets in two dimensional domains.

However, current research shows that the Jacobi Sets are unable to capture
the actual topological changes of multi-fields,
instead one should consider their Reeb Spaces,
introduced in \cite{2008-edels-reebspace}.
Recently, Multi-Dimensional Reeb Graphs \cite{2014-EuroVis-short} and
Layered Reeb Graphs \cite{2014-Strodthoff} have been introduced from
two different perspectives to extend the Reeb Graph for multi-fields. 
In the current paper, we use the recently introduced Jacobi
Structure \cite{2014-EuroVis-short} to separate the Reeb Space into regular and singular components.
Thus we obtain a dual Reeb Skeleton corresponding to the Reeb Space.
Our simplification strategy is based on simplifying this Reeb Skeleton
 by associating different measures with the nodes of the
Reeb Skeleton.
\section{Necessary Background}
\label{sec:Background}
\begin{figure*}[th!]
\begin{center}
\subfloat[\label{fig:1a}]{\includegraphics[height=5.5cm]{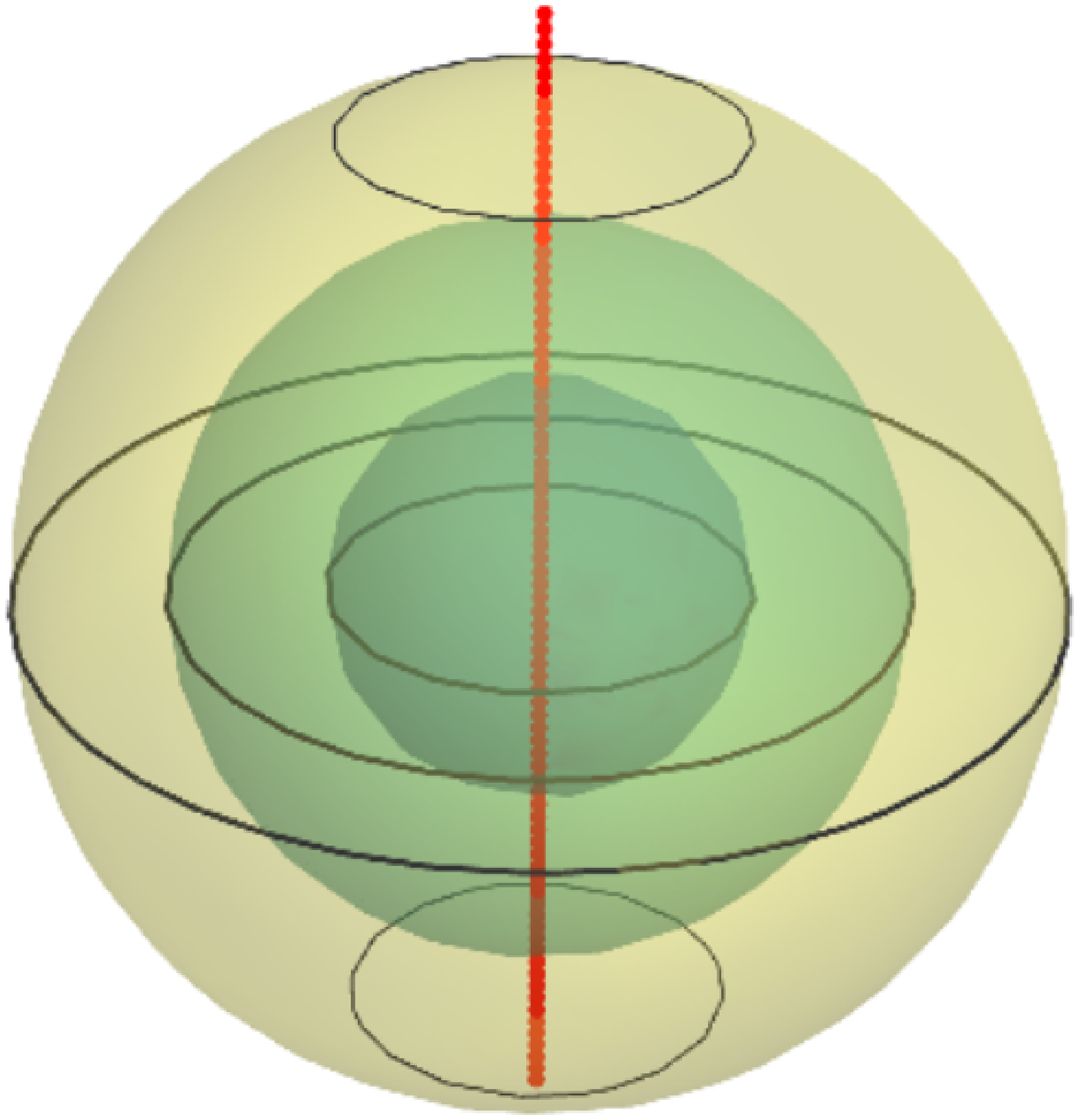}}
\subfloat[\label{fig:1b}]{\includegraphics[height=5.5cm]{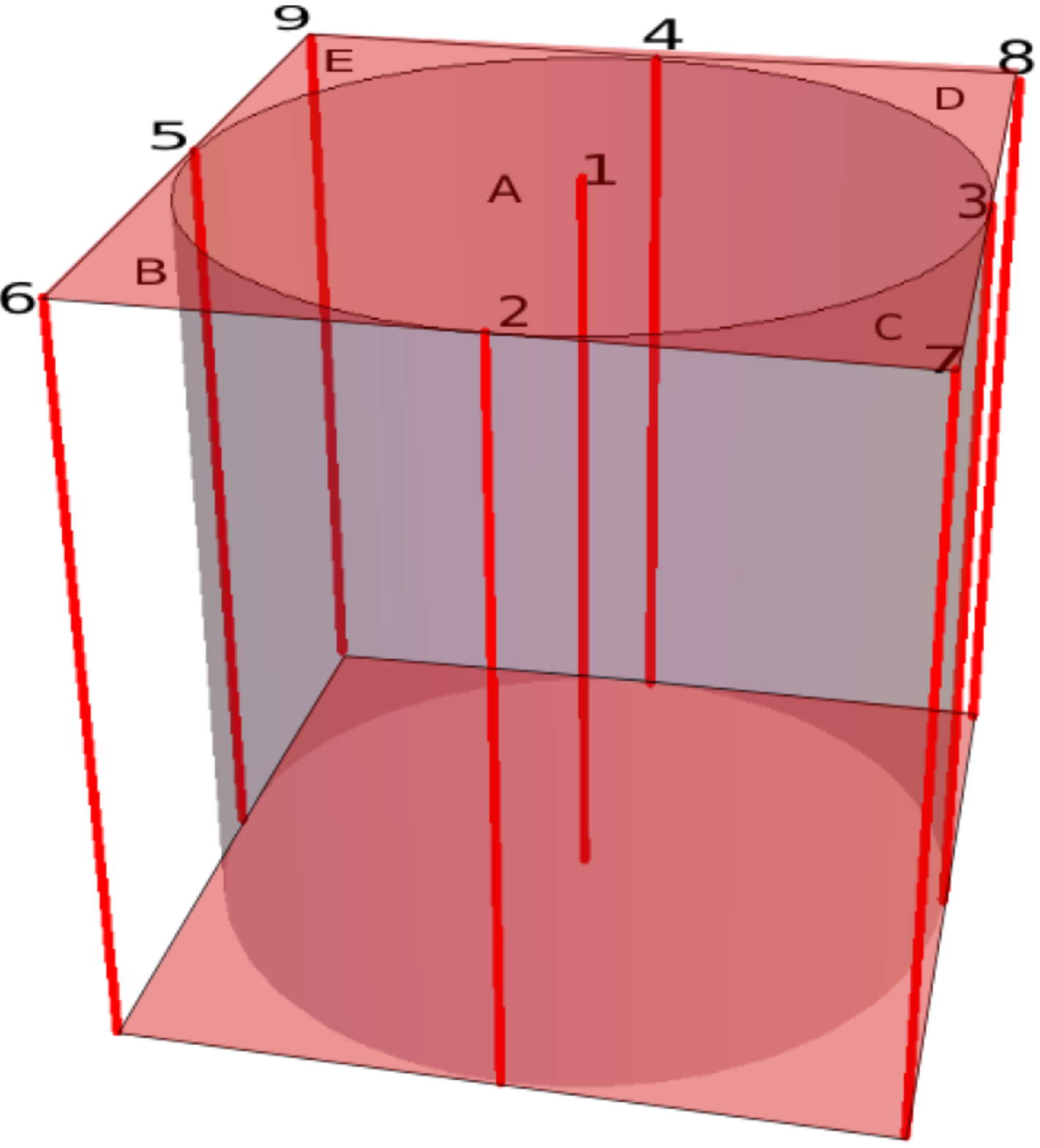}}\qquad \qquad
\subfloat[\label{fig:1c}]{\includegraphics[height=5.5cm]{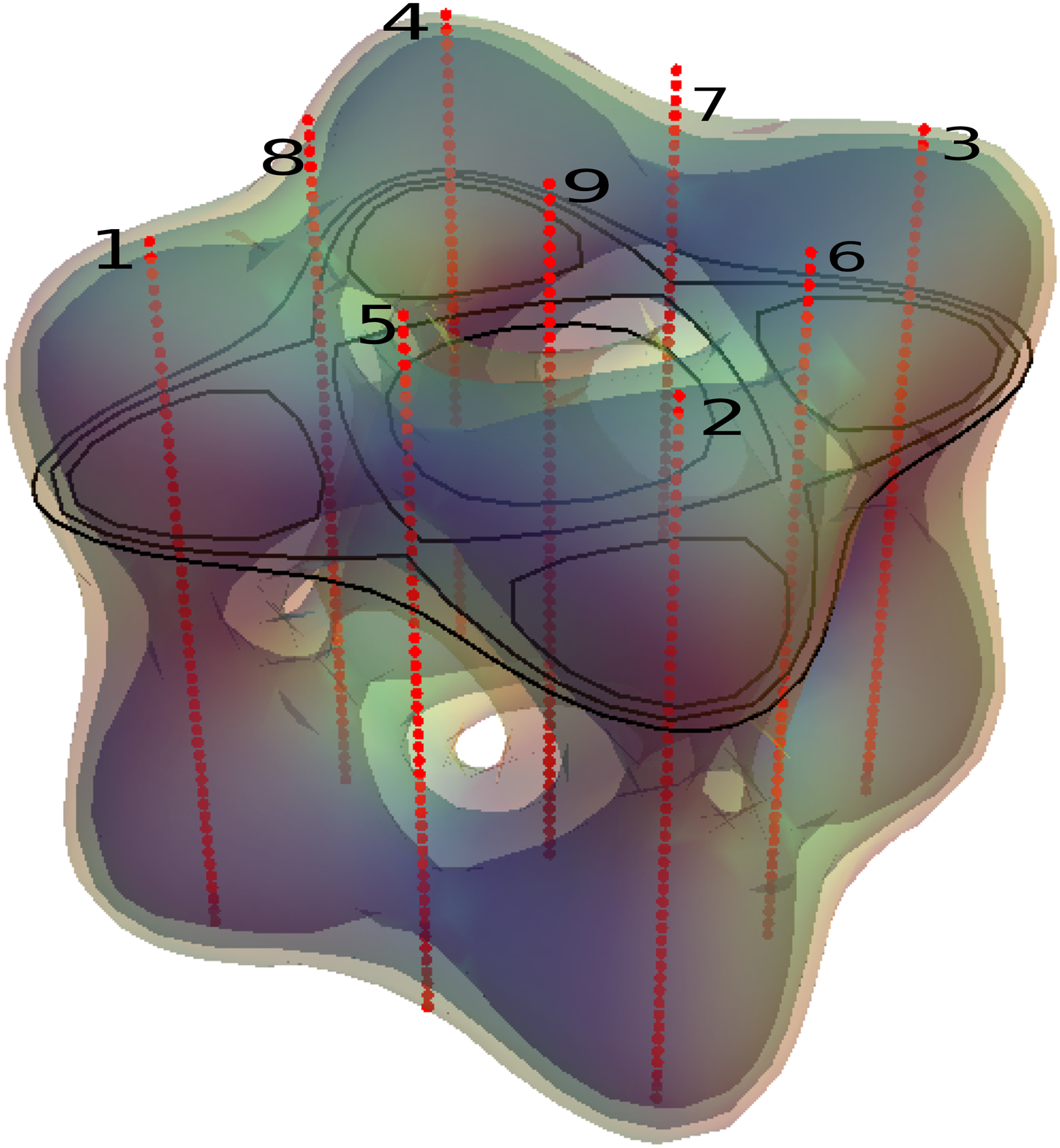}}

\subfloat[\label{fig:1d}]{\includegraphics[width=4.9cm]{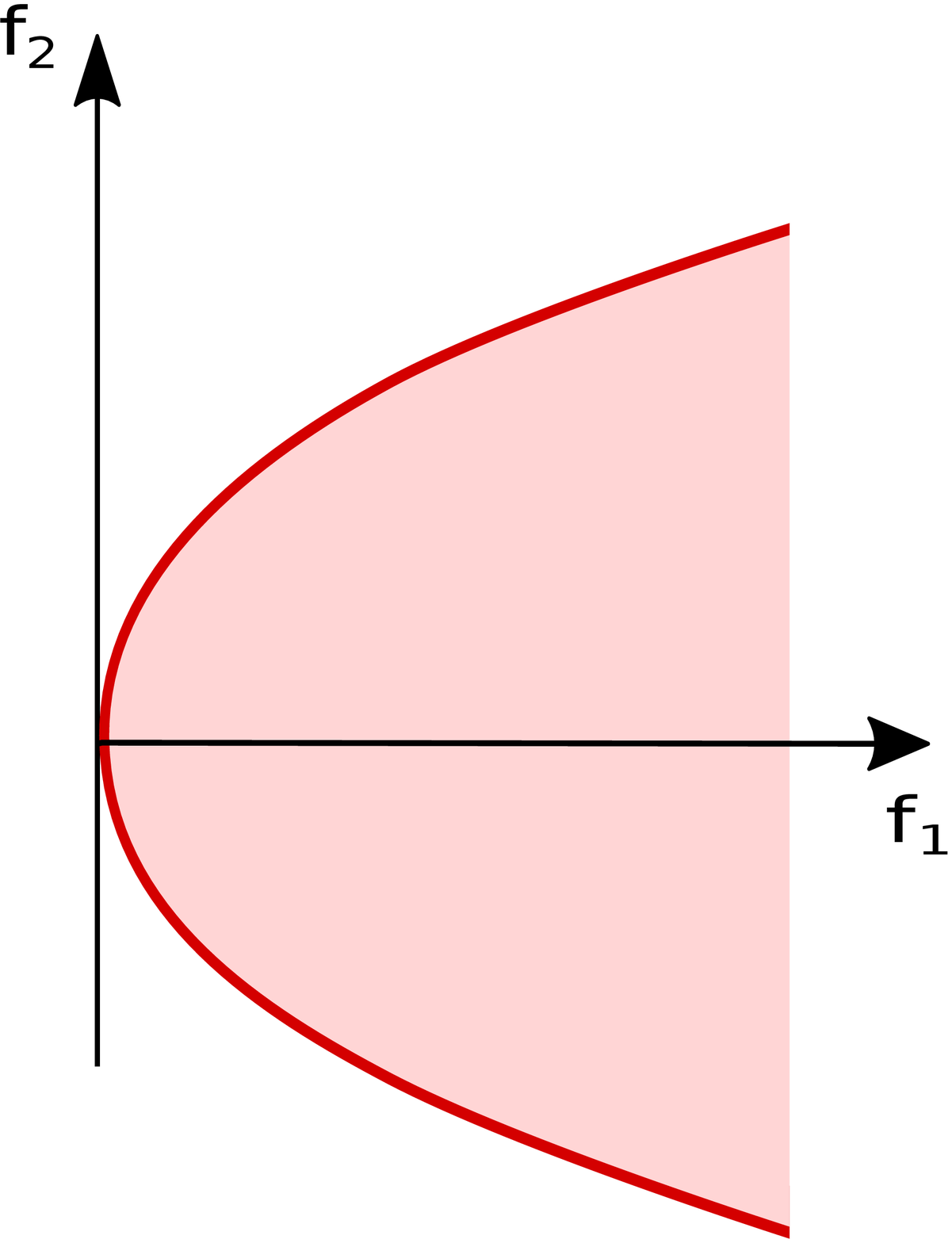}}\qquad \qquad
\subfloat[\label{fig:1e}]{\includegraphics[width=5.1cm]{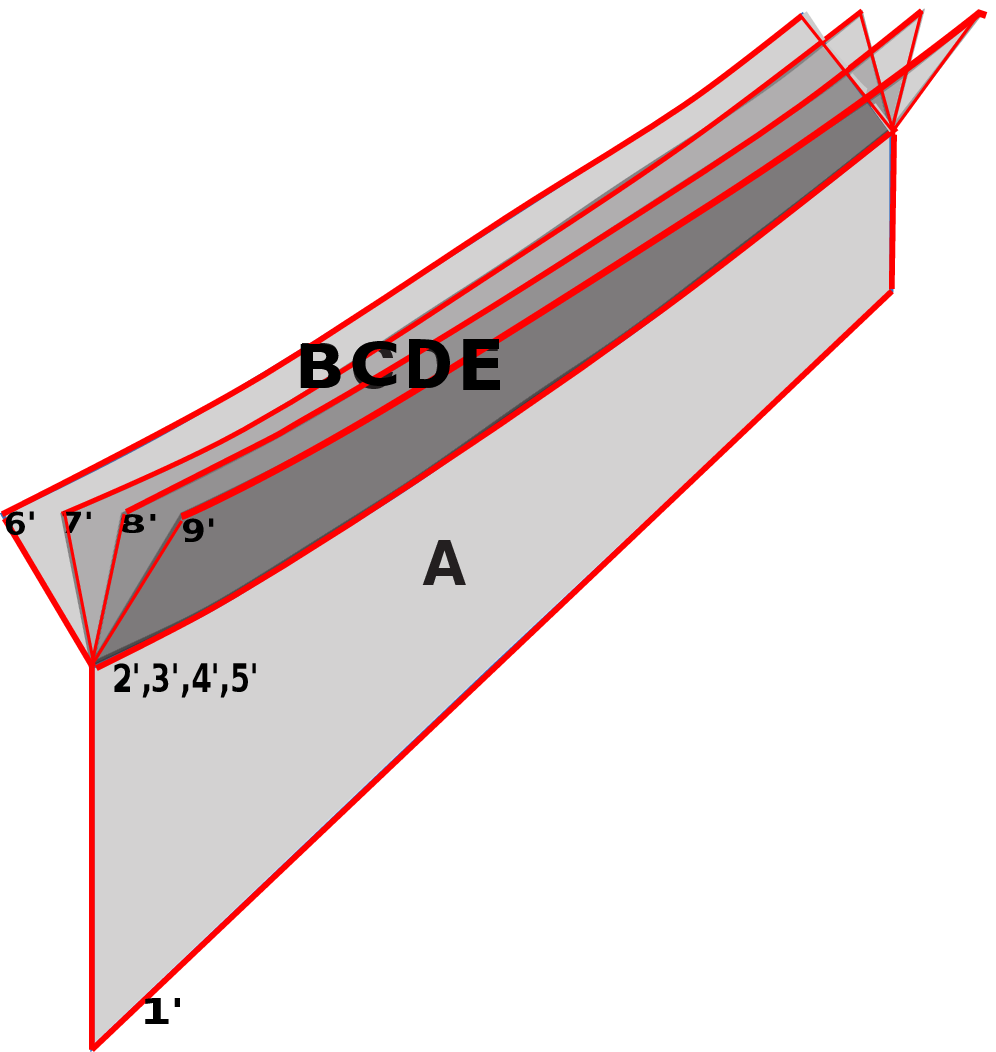}}\qquad \qquad
\subfloat[\label{fig:1f}]{\includegraphics[width=5.1cm]{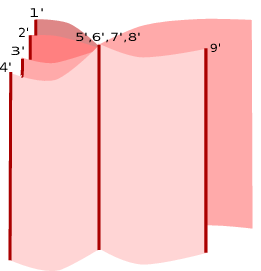}}
\end{center}
\vspace*{-2ex}
\caption{ (a) A stable bivariate field $(f_1, f_2)\equiv(x^2 + y^2 + z^2, \, z)$
  from $\R^3$ to $\R^2$ that is visualized using the transparent isosurfaces of the
  first component field; black curves are the fiber-components of the
  bivariate field;  the red line represents the Jacobi Set; 
(d) The Reeb Space corresponding to (a) that is comprising one sheet (in pink) and the Jacobi
  Structure (red parabolic curve); 
(b) The Jacobi Set (consists of the red
  lines, top face and bottom face of the box) of the bivariate field $(f_1, f_2)\equiv(x^2 + y^2 + z^2, \, z)$ in the
  box $[-1,\,1]\times [-1,\,1] \times[0,\,1]$; singular fibers passing through
  the boundary tangent points form a cylindrical surface that
  separates the domain into five components, denoted
  as A, B, C, D and E; 
(e) The Reeb Space of the multi-field
  corresponding to (b) that is comprising five sheets (in grey) and
  the Jacobi Structure (red lines); the regular
  components of the Reeb Space are marked to match the corresponding
  components in the  domain; components of the Jacobi Set in the domain and
  their corresponding projections in the Reeb Space are denoted by
  numbers; 
(c)  An unstable bivariate field $(f_1, f_2)\equiv (x^4 + y^4 + z^4 - 5(x^2 + y^2 +
  z^2) + 10, \, z)$ from $\R^3$ to $\R^2$ that is visualized using the transparent isosurfaces of the
  first component field; black curves are the fiber-components of the
  bivariate field;  the Jacobi Set consists of 9 red lines; 
(f) The Reeb Space corresponding to (c) that is comprising six sheets
and the Jacobi Structure (6 red lines).}
\label{fig:jacobi-reeb}
\end{figure*} 

Over the last two decades, scalar topology has been used to support scientific data analysis and visualization,
in particular through the use of the Reeb Graph and its
specialisation, the contour tree \cite{1997-Kreveld-CT,
  2000-Carr-CT, 2001-Hilaga, 2004-edels-timevarying, 2007-Pascucci}.
The subject of multi-field topology in data analysis is
rather new. In this section we briefly describe the multi-field
topological analysis and existing tools for capturing them, viz. the
Jacobi Set and the Reeb Space.

\begin{table}[h!]
 \begin{center}
 \caption{Important Notations}
\begin{tabular}{ll}
\hline
\textbf{Notation}        &  \textbf{Name} \\ 
\hline 
$\RS$ & 	          Reeb Space\\
$\RK$ &            Reeb Skeleton\\
$\Jf$  &             Jacobi Set\\
$\Jb$ &             Boundary Jacobi Set\\
$\Ji$  &             Interior Jacobi Set \\
$\JS$ &             Jacobi Structure\\
$\JCN(f, m_Q)$ &          Joint Contour Net with quantization
              level $m_Q$\\
$\RG$ &           Reeb Graph\\
$\MDRG$&       Multi-dimensional Reeb Graph\\
\hline
\end{tabular}
\label{tab:notations}
\end{center}
\end{table}

\subsection*{Multi-Field Analysis.}
A multi-field on a $d$-manifold $\mathbb{X}\, (\subseteq\mathbb{R}^d)$ with $r$ component scalar fields
$f_i:\mathbb{X}\rightarrow \mathbb{R}$ ($i=1,\,\ldots, r$) is a \textit{map} $f=(f_1,\,f_2,\,\ldots,\,f_r): \mathbb{X}\rightarrow
\mathbb{R}^r$. Table~\ref{tab:notations} shows the notations used to denote
various structures corresponding to a multi-field $f$ in the current paper.

In differential topology, $f$ is considered to be a \textit{smooth map} when all its partial derivatives of any order
are continuous. A point $\bx\in \mathbb{X}$ is called a \textit{singular point} (or
\textit{critical point}) of $f$ if the
rank of its differential map $df_{\bx}$ is strictly less than $\min\{d,
r\}$ where $df_{\bx}$ is the $r \times d$ matrix whose rows are the
gradients of $f_1$ to $f_r$ at $\bx$. And the corresponding
value $f(\bx)=\bc=(c_1,\, c_2,\, \ldots,\, c_r)$ in $\mathbb{R}^r$ is a \textit{singular
value}. Otherwise if the rank of the differential map $df_{\bx}$ is $\min\{d,
r\}$ then $\bx$  is called a \textit{regular point} and a point $\mathbf{y}\in \R^r$
 is a \textit{regular value} if $f^{-1}(\mathbf{y})$ does not contain
 a singular point.

The inverse image of the map $f$ corresponding to a
value $\bc\in \mathbb{R}^r$, $f^{-1}(\bc)$ is called a \textit{fiber}  and each
connected component of the fiber is called a \emph{fiber-component}
\cite{Saeki2014, 2004-Saeki}. In particular, for a scalar field these are known as
the \emph{level set} and the \emph{contour}, respectively.
The inverse image of a singular value is called a \textit{singular fiber} and the inverse image
of a regular value is called a \textit{regular fiber}. If a
fiber-component passes through a singular point, it is called a \emph{singular
fiber-component}. Otherwise, it is known as a \emph{regular
fiber-component}. Note that a singular fiber may contain a regular fiber-component.

A continuous map is said to be \emph{proper} if the pre-image
of a compact set is always compact and  it is said to be \emph{stable} if its topological properties remain unchanged
by small perturbations \cite{Levine1985}. 
Let $f : \X \subset \R^3 \to \R^2$ be a proper smooth map. Then, it is stable if and only if it satisfies the following local and global conditions.
Around each singular
point $\bs$, $f$ is locally described as either (i) $(u,\,
x^2+y^2)$: $\bs$ is a definite fold point, or (ii) $(u,\, x^2-y^2)$: $\bs$ is an indefinite fold point, or (iii) $(u,\,
y^2+ux-x^3/3)$: $\bs$ is a cusp point, for some local coordinates $(u,\, x,\, y)$  around $\bs$
and an appropriate set of local coordinates around $f(\bs)$ in the range $\R^2$.
Moreover, no cusp point is a double point of $f$ restricted to the set
of singular points and $f$ restricted to the set of all definite and
indefinite fold points is an immersion with normal crossings. 
Thus for a proper stable map, a singular fiber-component passing through a definite fold
point or a cusp point contains exactly one such point, while
a singular fiber-component passing through an indefinite fold point may pass through one or two indefinite fold
points. Otherwise, the map is called an \emph{unstable map}.

We note, characterizing the stability of the maps on compact 3-manifold domains with boundary 
needs additional types of singularities which is discussed in \cite{Saeki2015}.
\figref{1a} is an example of a map from $\R^3$ to $\R^2$ where $f=(x^2+y^2+z^2,\,z)$.
All its singular points are definite fold points, so this is an example of a stable map. \figref{1c}
is an example of a map from $\R^3$ to $\R^2$ where $f=(x^4 + y^4 + z^4 - 5(x^2 + y^2 +
  z^2) + 10, \, z)$. It has singular fiber-components which pass
  through four indefinite fold points (on corresponding four 1-manifold components
  numbered as 5, 6, 7 and 8 in \figref{1c}) and so is an example of an unstable map.

From the pre-image theorem \cite{1974-gp-dt}, generically a regular fiber
$f^{-1}(\bc)$ is a $(d-r)$-manifold for the regular value $\bc=(c_1,\, c_2,\, \ldots,\, c_r)$. 
We note for $d<r$, $f^{-1}(\bc)$ is an empty set or a discrete set of points.
A fiber $f^{-1}(\bc)$ can be considered as the intersection of
the fibers of the component scalar fields
$f_1^{-1}(c_1),\, f_2^{-1}(c_2),\ldots ,\, f_r^{-1}(c_r)$ and
a connected component of this intersection is a
fiber-component. Alternatively, fiber-components of
$(f_1,\,f_2,\,\ldots,\,f_r)$ can be considered as the contours of a
component field $f_i$, restricted to the fiber-components of the
remaining component fields. This is a \textit{key observation}, we use
in building our Multi-Dimensional Reeb Graph data-structure.

\subsection*{Jacobi Set.}  The compact $d$-manifold domain $\X \, (\subseteq\mathbb{R}^d)$ of the map $f$ can be expressed as
$\mathbb{X}=\Int{\X}\cup \partial\X$ where
$\Int{\X}$ denotes the interior (the set of interior points) of
the domain  and $\partial\X$ denotes the boundary (the set of boundary
points) of $\X$.  In case the domain $\X$ is without boundary, $\partial
\X=\emptyset$ and $\X=\Int{\X}$. 

Now the \emph{interior Jacobi Set} of  the map
$f:\mathbb{X}\rightarrow\mathbb{R}^r$
is denoted by $\Ji$ and is defined by the set 
$\Ji := \left\{ \bx \in \Int{\X} \mid \text{ rank }
  d\Int{f}_x<\min\{d, r\} \right\}$ \cite{2004-edels-jacobi} where
$\Int{f}$ is the restriction of $f$ to $\Int{\X}$, i.e., $\Int{f}:= f|_{\Int{\X}}:{\Int{\X}}\rightarrow \mathbb{R}^r$.
In other words, $\Ji$ is the set of singular points of the map
$f$ interior to the domain $\X$. Similarly, the \emph{boundary Jacobi
  Set} of the map $f$ is denoted by $\Jb$ and is defined as the set of singular points of the
restriction of $f$ to the boundary $\partial\X$, i.e.  $f_\partial:=f|_{\partial  \X}: \partial \X\rightarrow \mathbb{R}^r$. 
Finally, by the \emph{Jacobi Set} of the map $f: \X\rightarrow\R^r$ we mean
the union of the interior and the boundary Jacobi Set of the map $f$,
and is denoted by $\Jf$, i.e. $\Jf=\Ji\cup \Jb$. 

Now the boundary of a domain may come with \emph{corners}, e.g. a
3-dimensional cube has corners
of two types: 12 edge corners, and 8 vertex corners (as in \figref{1b}). Let $\X$ be a compact $3$-dimensional manifold
with corners and $f : \X \to \R^2$ be a smooth
map. A point $q \in \partial \X$ is a (boundary)
regular point if $f$ restricted to $\partial \X$
is a local homeomorphism around $q$, where
$\partial \X$ stands for the boundary of $\X$ which
includes all the boundary points and corner points.
Otherwise, $q$ is a (boundary) singular point.
For example, if we take a point in a vertical edge in \figref{1b},
then in its (2-dimensional) neighborhood on
the boundary, there are always a pair of points
that are mapped to the same point.
Thus, it is never injective, and hence is never a
local homeomorphism. Therefore, the point
is a (boundary) singular point.

Alternatively, the Jacobi Set is the set of  critical
points of one component field (say $f_i$) of $f$ restricted to the intersection of the level
sets of the remaining component fields. Edelsbrunner et al. \cite{2004-edels-jacobi} studied properties of the Jacobi Set for $r$
Morse functions. They proved the Jacobi Set is symmetric with respect to its component
fields. They also showed, generically, the Jacobi Set of two Morse functions is a smoothly embedded
$1$-manifold where the gradients of the functions become
parallel. However, in general Jacobi Sets
are not sub-manifolds of the domain of the multi-field $f$,
and are the disjoint union of sub-manifolds
of the domain \cite{2004-edels-jacobi}. 
The red lines in \twofigref{1a}{1c}
illustrate the Jacobi Sets of multi-fields on domains without boundary.

\subsection*{Reeb Space.} As with the Reeb Graph of a scalar field,
 the Reeb Space parametrizes the fiber-components of a
 multi-field and its topology is described by the
 standard quotient space topology.
We note the fiber-components of a continuous map
$f:\mathbb{X}\rightarrow \mathbb{R}^r$ ($\mathbb{X}\subseteq
\mathbb{R}^d$) partition the domain $\mathbb{X}$ into a set 
of equivalence classes, denoted by $\RS:=\mathbb{X}/\sim$,
where two points $a,\,b\in \mathbb{X}$ are equivalent or $a\sim b$ if
$f(a)=f(b)$ and $a,\, b$ belong to the same fiber-component of $f^{-1}(f(a))$ and $f^{-1}(f(b))$. 
Now the canonical projection map
$q_f:\mathbb{X}\rightarrow \mathbb{X}/\sim$ that maps each element of $\mathbb{X}$
to its equivalence class defines the standard quotient topology where
open sets are defined to be those sets of equivalence classes with an
open pre-image, under map $q_f$. The Reeb Space of $f$ is the quotient
space $\RS$ together with this quotient topology.
The decomposition of $f$ as the composition of $q_f$ and $\bar{f}$,
where $\bar{f}:\RS\rightarrow\R^r$ is such that $f=\bar{f}\circ q_f$. This is called the Stein factorisation of $f$.
The following commutative diagram describes this relationship between the maps.
\begin{center}
\begin{tikzcd}[column sep=normal]
\mathbb{X} \arrow{dr}[swap]{q_f}\arrow{rr}{f} & & \mathbb{R}^r\\
& \RS \arrow{ur}[swap]{\bar{f}} &
\end{tikzcd}
\end{center}
\noindent
Now to construct a fiber $f^{-1}(a)$, instead of going directly from
$\mathbb{R}^r$ to $\mathbb{X}$ one can compute the pre-image under $\bar{f}$ of
$a$. Each fiber consists of a number of components, one for each point in $\bar{f}^{-1}(a)$.
Generically, the Reeb Space is a Hausdorff space, i.e., any two distinct points of
 $\RS$ have disjoint neighbourhoods.
Moreover, when $r\leq d$ the Reeb Space corresponding to the multi-field
$f$ consists of a collection of $r$-manifolds glued together in
complicated ways \cite{2008-edels-reebspace}.

\threefigref{1d}{1e}{1f} show three examples of the Reeb Spaces
corresponding to a stable bivariate
field in $\R^3$, an unstable bivariate field on a closed 3-dimensional interval
and an unstable bivariate-field in $\R^3$, 
respectively. We indicate the dark (red) lines in the Reeb Spaces as the Jacobi Structures which are introduced in the next section.
Note that the structures of
the Reeb Spaces as in \threefigref{1d}{1e}{1f} are obtained by analyzing the evolution of the
fiber-components of the corresponding bivariate fields. For example, if we consider evolution
of the fiber-components of the map in \figref{1b}, 
they start at the definite fold points on the line numbered as 1.
Then these fiber-components start growing and meet at the boundary
Jacobi Set points (on the lines numbered as 2, 3, 4, 5). 
Then each of them splits into four fiber-components which continue to shrink 
and die at the corner singular points (on the lines numbered as 6, 7,
8, 9). This evolution phenomenon is captured in its Reeb Space (e). 
\section{Theoretical Results}
\label{sec:Theory}
In this section we exploit the underlying structure of the Reeb Space
to decompose it into a set of simple manifold-like components (namely
regular and singular components) and capture 
their connectivities by a dual skeleton graph (namely Reeb Skeleton).
Then, since in real applications most of the data come with simple
domains (such as, a cube or a box), we study properties of the Reeb Space and its representative skeleton graph
for such topologically simple data-domains. More precisely, to prove our 
theoretical results in this section we consider a stable bivariate
field $f=(f_1,f_2):\X\rightarrow\R^2$ where $\X\; (\subseteq\R^3)$ is a
three-dimensional bounded, closed interval. 
However, most of the results
are straight-forward to generalize for multi-fields of  
higher dimensions. Thus the domain $\X$, we consider, is a compact domain with
boundary and is simply-connected and in this case 
the Reeb Space is path-connected.

\subsection{Path Connectedness}
For a continuous map $f: \mathbb{X}\subseteq \mathbb{R}^3\rightarrow
\mathbb{R}^2$ the Reeb Space is a quotient space of the
fiber-components and is path-connected (or $0$-connected). That is, any two points $p_0$ and $p_1$ of the
Reeb Space can be connected by a path $\gamma: [0,1]\rightarrow \RS$
so that $\gamma(0)=p_0$ and $\gamma(1)=p_1$. In other wards, we say $0$-connectivity is
preserved by the quotient map $q_f: \mathbb{X} \rightarrow \RS$. This can also be stated by
saying that $0$-th homotopy group of the Reeb Space $\pi_0(\RS)$ remains trivial.
Next we prove the following important property of the Reeb Space.

\begin{lemma}
\label{lem:path}
Let $f:\mathbb{X}\subseteq \mathbb{R}^3\rightarrow \mathbb{R}^2$ be a continuous, generic map on a $3$-dimensional interval $\mathbb{X}$  and
$\RS$ be the corresponding Reeb Space. Let $P$ be a
continuous path between any two points on the Reeb Space. Then
if $\RS \setminus P$ is path-connected, then so
is $\mathbb{X}\setminus q_f^{-1}(P)$.
\end{lemma}
\noindent
\begin{proof}
Consider any two points $p_0,\,p_1\in \RS \setminus
P$. Since $\RS \setminus P$ is path-connected, $\exists$ a path
$\gamma: [0, 1]\rightarrow \RS \setminus P$ with $\gamma(0)=p_0$ and
$\gamma(1)=p_1$.  Using conditions like
the genericity of $f$, $\gamma(t)$ lifts to $\X \setminus q_f^{-1}(P)$,
i.e., there exists a path $\tilde{\gamma}$ in
$\X$ such that $q_f \circ \tilde{\gamma}(t) = \gamma(t)$ (using the
\emph{path-lifting} property \cite{2002-Hatcher}).
Now $\tilde{\gamma}$ is a path between any point of $q_f^{-1}(p_0)$ to any point
of $q_f^{-1}(p_1)$ in $\mathbb{X}\setminus q_f^{-1}(P)$. Therefore, $\mathbb{X}\setminus q_f^{-1}(P)$ must be path-connected.
\end{proof}

\noindent
Thus Lemma~\ref{lem:path} implies if there
exists a path $P$ in the Reeb Space whose preimage $q_f^{-1}(P)$ separates
the domain then $P$ must also separate the Reeb Space. This is a 
useful property in detaching unimportant components from the Reeb Space.

\subsection{Jacobi Structure}
\label{subsec:Jacobi}
As noted in Section \ref{sec:Background}, the Jacobi Set of a function is not the same as the set of singular fibers, as each 
point in the Jacobi Set is merely a representative of a singular fiber. Moreover, the structure of 
the Reeb Space is actually given by a projection of the Jacobi Set or 
the singular fibers. For example, in 
\twofigref{1c}{1f}, the Jacobi Set consists of 9 parallel lines in the domain, but
they correspond to 6 1-manifold structures in the Reeb Space. 
Note that if the input multi-field domain is with boundary, there are
additional edges (corresponding to the boundary
 Jacobi Set) needed to describe the Reeb Space.
We therefore introduce the \emph{Jacobi Structure}: the 
manifold structure of the Reeb Space corresponding to the Jacobi Set
in the domain.

\begin{dfn}
The \textbf{Jacobi Structure} of a Reeb Space $\RS$ corresponding to a
multi-field $f:\mathbb{X}\subseteq \mathbb{R}^3\rightarrow \mathbb{R}^2$ is denoted
by $\JS$ and is defined by $\JS:=q_f(\Jf)$, i.e., the projection of the Jacobi Set
$\Jf$ to the Reeb Space by the quotient map $q_f:\mathbb{X}\rightarrow\RS$.
\end{dfn}

\noindent
Note that according to our definition the Jacobi Set $\Jf$ consists of both the interior and the
boundary Jacobi Set, i.e., $\Jf=\Ji\cup \Jb$.
Thus each point of the Jacobi Structure corresponds to a singular fiber-component
in the domain $\X$ of $f$, and vice-versa.

\begin{figure}[t!]
\begin{center}
\includegraphics[width=.47\textwidth]{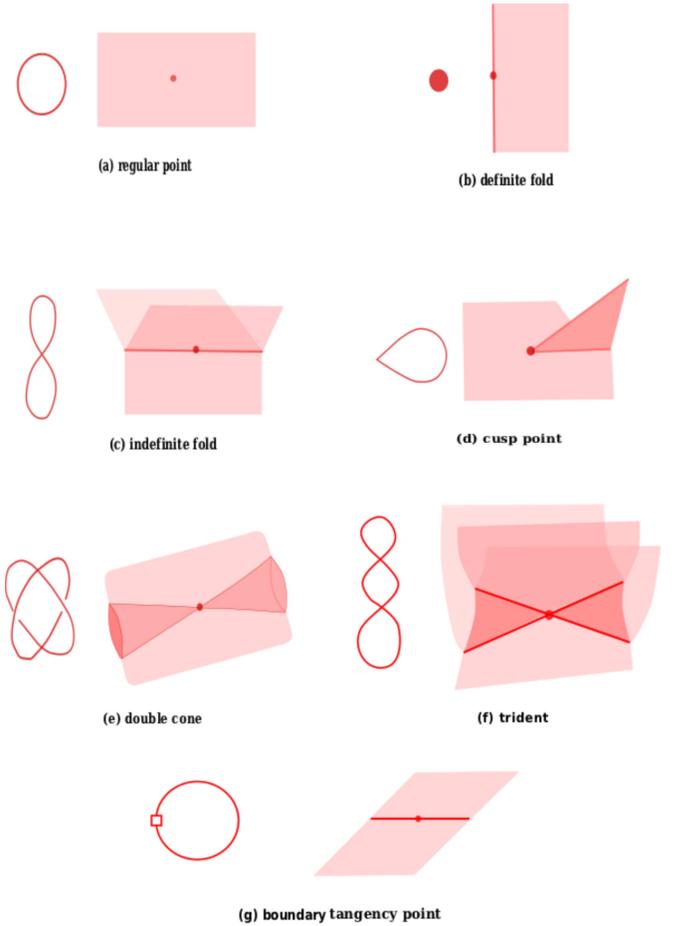}
\end{center}
\caption{Regular and singular fiber-components and corresponding local configurations in
  the Reeb Space. The Jacobi Structures are in red lines in the Reeb
  Space \cite{Saeki2015}.}
\label{fig:jacobi}
\end{figure}

To understand the underlying structure of the Reeb Space
one needs to understand both the topology of the singular fibers 
and the corresponding local configurations of the Jacobi Structure. Classification of singular
fibers and their local configurations in the quotient space have been studied
for stable maps from $\mathbb{R}^3$ to $\R^2$ and $\R^4$ to $\R^3$~\cite{2004-Saeki}.
Figure \ref{fig:jacobi} illustrates examples of a regular and few
singular fiber-components, and their local structures in the Reeb Space
\cite{Kushner-1984, Levine1985} for a stable map $f:\mathbb{X}\subseteq
\mathbb{R}^3\rightarrow \mathbb{R}^2$.
For a more complete classification of singular fibers for maps on 
3-manifolds (with boundary) to plane and for local configurations of the Reeb spaces we refer to \cite{Saeki2015, SaekiCobordism2015}.

For a generic map $f:\X\rightarrow\R^2$, $\RS$ is a two-dimensional
polyhedron and Jacobi structure embedded in the Reeb Space consists of
1-dimensional components which are at the boundary of the
two-dimensional sheets in $\RS$.
Now a 1-manifold component of the Jacobi Structure can be classified into three types
based on  the transition of number of regular fiber-components if one
passes across the component \cite{Saeki2014}:

\begin{enumerate}
\item \emph{Birth-Death} or \emph{Boundary component} - where a fiber-component takes
birth or dies (Figure~\ref{fig:jacobi} (b)), 
\item \emph{Merge-Split component} or \emph{Bifurcation locus} - where
  two (or more) fiber-components merge together or one component splits into two (or more) (Figure~\ref{fig:jacobi}(c))
and 
\item \emph{Neutral component} - where there is no
change in the number of fiber-components if one passes through such
components (Figure~\ref{fig:jacobi}(g)), but here, the topology of the regular fiber-component changes from a circle to an arc (or vice versa).
\end{enumerate}
A connected component of the Jacobi structure may also consist of a composition of these
three types, e.g. in Figure~\ref{fig:jacobi}(d) the Jacobi structure
component consists of a boundary and a merge-split
component connected at a discrete \emph{cusp point}. In
Figure~\ref{fig:jacobi}(e) four merge-split components are connected
at a \emph{double point} on the Jacobi Structure.

\twofigref{1b}{1e} respectively show an example of 8 1-manifold components
of the boundary Jacobi Set (red lines in the boundary of the domain) and their
corresponding projection in the Reeb Space as 5 1-manifold parts of
the  Jacobi Structure. From this example, it is clear
that a boundary Jacobi Set component may not be the boundary
component of the Jacobi Structure in the Reeb Space or vice-versa. 
In Section~\ref{sec:SimplifyingJCN} we propose an algorithm for computing the Jacobi Structure
by constructing a Multi-Dimensional Reeb Graph corresponding to a multi-field.

\subsection{Regular and Singular Components}
As the number of dimensions increases, the projections of the singular fibers
develop more internal structure in the Reeb Space.
Consider the Reeb Graph of a scalar function: in this, the projection
images of the critical points are single points (0-manifolds)
separating edges (1-manifolds). Similarly, for the bivariate fields
shown in \threefigref{1d}{1e}{1f}, the projections of the
singular fibers are arranged in a Reeb Space along 1-manifold curves which separate 2-manifold sheets.
This induces a natural stratification or partition of the Reeb Space
into disjoint subspaces (or strata).

To describe a stratification of the Reeb Space and the corresponding
domain of the multi-field we first classify the fiber-components of the generic map $f: \X\subseteq
  \R^3\rightarrow\R^2$ according to their complexity or codimension of the subspace where they lie \cite{Saeki2014}.
 Given the Stein factorization $f=\bar{f}\circ q_f$, fiber-components  of $f$ can be classified into three classes.
\begin{enumerate}
\item $\mathcal{C}^0=\{q_f^{-1}(s): s\in \RS \text{ and } q_f^{-1}(s) \text{ does
  not contain any }$ $\text{singular point of } f\}$. Fiber-components
of this class are the regular fiber-components and their $q_f$-images 
form codimension 0 subspaces in $\RS$, denoted as $\RS^0$. 

\item $\mathcal{C}^1=\{q_f^{-1}(s): s\in \RS \text{ and } q_f^{-1}(s)
  \text{ contains exactly one de-}$ $\text{finite or indefinite fold point}\}.$
Singular fiber-components of this class are moderately complex and
  their $q_f$-images form codimension 1 subspaces  in $\RS$, denoted as $\RS^1$.

\item $\mathcal{C}^2=\{q_f^{-1}(s):  s\in \RS \text{ and } q_f^{-1}(s)
  \text{ contains a cusp point}$ $\text{or two indefinite fold points}\}.$ 
 Singular fiber-components of this class
  are the most complex and their $q_f$-images form codimension 2
  subspaces  in $\RS$, denoted as $\RS^2$.
\end{enumerate}

\noindent
Complexity of a fiber-component increases as the codimension of the
corresponding subspace in the Reeb Space increases.
Note that $q_f$-images of the fiber-components in $\mathcal{C}^1$ and $\mathcal{C}^2$ form the Jacobi
Structure $\JS$ of the Reeb Space, i.e., $\JS=\RS^1\cup \RS^2$. Topologically, regular fiber-components are
either a circle or an arc \cite{Saeki2015}. For stable maps $f: \X\subseteq
  \R^3\rightarrow\R^2$, topologically there are $7$ different types of singular
  fibers in $\mathcal{C}^1$ and $21$ different types of singular
  fibers in $\mathcal{C}^2$ \cite{Saeki2015}.

Two regular points $a,\, b \in \RS^0$ are \emph{topologically
equivalent} in the Reeb Space $\RS$ or $a\sim_\rho b$ if there exists a path
between $a$ and $b$ without intersecting the Jacobi Structure $\JS$.
It is not difficult to check that `$\sim_\rho$' is an equivalence relation.
Therefore, the equivalence relation `$\sim_\rho$' partitions the regular points
of $\RS$ into a set of equivalence classes.
Now we prove that each such equivalence class is a 2-dimensional sheet.

\begin{lemma}[\textbf{Partition}]
\label{lem:partition}
The Jacobi structure $\JS$  of a Reeb space $\RS$ corresponding to a smooth
stable map $f:\mathbb{X}\subseteq \mathbb{R}^3\rightarrow \mathbb{R}^2$ separates  the Reeb Space into a set
of $2$-manifold components.
\end{lemma}
\noindent
\begin{proof}
Let $D$ be a small disk in the range consisting of
regular values (i.e., $D$ does not intersect $f(\J_f)$).
Then, by Ehresmann's fibration theorem,
$f$ restricted to $ f^{-1}(D)$ is equivalent to
the projection $D \times F \rightarrow D$, where
$F$ is a 1-dimensional compact manifold.
So, this means that $q_f(f^{-1}(D))$ can
be identified with a disjoint union
of some copies of $D$, where the number of copies
is the same as the number of connected components
of $F$. Even when $D$ intersects with $f(\J_f)$,
if we restrict $f$ to the components
of the inverse image $f^{-1}(D)$ that do not
intersect $\J_f$, then the same consequence holds.
So, the regular sheets of $\RS$ are locally
homeomorphic to $D$, and hence is a 2-manifold.
\end{proof}

\noindent
Thus we have the following definition of regular components.

\begin{dfn}
A path-connected component of $\RS\setminus\JS$ or $\RS^0$ is called a regular component. 
\end{dfn}

\noindent
Generically, the $0$-dimensional strata are in the boundary of
the $1$-dimensional strata in the $\JS$.
Therefore, an equivalence relation on the set of points in $\RS^1$  can
be defined, similarly, where two points of $\RS^1$ are equivalent if there exists a continuous path between
them without crossing the $0$-dimensional strata in $\JS$ and each such equivalence
class will be considered as a 1-singular component.

\begin{dfn}
A path-connected component of $\JS\setminus \RS^2$ or $\RS^1$ is called a 1-singular component.
\end{dfn}

\noindent
Note that a 1-singular component in $\RS$ may be an arc or a circle. An arc
1-singular component will also be called as an \emph{edge}.

\begin{dfn}
Each component of $\RS^2$ is called a 0-singular component.
\end{dfn}

\noindent
To extract a skeleton graph from the Reeb Space we need adjacency of
these regular and 1-singular components which are defined as follows.

\begin{dfn}
\begin{enumerate}
\item A circle 1-singular component is self-adjacent (adjacent to itself).
\item If two end points of an arc 1-singular component coincide, then the
1-singular component is self-adjacent.
\item Two distinct 1-singular components $S_1,\,S_2$ are
adjacent if $\exists$ a 0-singular component $\alpha_0$
such that $S_1~\cup~ S_2~\cup~\alpha_0$ form a connected space.
\end{enumerate}
\end{dfn}

\begin{dfn}
A 1-singular component $S_i$ is adjacent to a
regular component $R_j$ if $S_i\cup R_j$ forms a connected space.
\end{dfn}

Next we define a connectivity graph of regular and singular components
based on their adjacency.

\begin{figure}[t!]
\begin{center}
\fbox{\includegraphics[width=.47\textwidth]{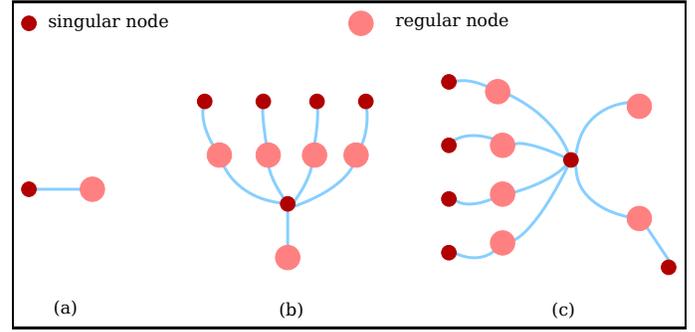}}
\end{center}
\vspace*{-2ex}
\caption{Reeb Skeletons: (a) corresponding to the Reeb Space in \figref{1d}, (b)
  corresponding to the Reeb Space in \figref{1e}, (c) corresponding to the
  Reeb Space in \figref{1f}.}
\label{fig:reeb-skeleton}
\end{figure}
\subsection{Reeb Skeleton}
\label{sec:ReebSkeleton}
Once the Reeb space $\RS$ is split into $2$-manifold regular components
and $1$-or-lower manifold singular components, it is possible to
perform a further reduction from the Reeb Space. To do so, we
represent both these regular and 1-singular components as points (or nodes), and add edges representing their
adjacency: in short, we can build the dual graph of these components of the Reeb Space.  This has
the merit of further reducing the Reeb Space from a $2$-dimensional structure to a fundamentally 
$1$-dimensional structure which is easier to represent, to reason about and to visualise.  We refer to this
as the \emph{Reeb Skeleton} and formally define as follows. 

\begin{dfn}
\label{dfn:reeb-skeleton}
 Let $R_1, R_2, \ldots, R_m$ be the regular components and $S_1, S_2,
 \ldots, S_n$ be the 1-singular components of $\RS$. Then
the Reeb Skeleton of $f$, denoted by $\RK$, is the adjacency graph which consists of
(i) nodes  $n_{R_i}$ and $n_{S_j}$ ($i=1, 2, \ldots, m$ and $j=1, 2,
\ldots, n$) corresponding to each of the regular and 1-singular
components,  and (ii) edges $e(S_j, S_{j'})$ and $e(R_i, S_j)$ that are
defined as follows:
\begin{enumerate}
\item If $S_j$ is self-adjacent, then $e(S_j, S_j)=1$. In other words, $n_{S_j}$ has
a self-loop.
\item If $S_j$ is self-adjacent and $S_j$ is adjacent with a regular component $R_i$, then
$e(R_i, S_j)=2$. In other words, $n_{S_j}$ is connected with $n_{R_i}$ by two edges.
\item If $S_j$ and $S_{j'}$ are two distinct \textbf{non-boundary}
  1-singular components, then
\begin{align*}
e(S_j, S_{j'})=\left\{
\begin{array}{ccc}
1, & \text{ if } S_j,\, S_{j'} \text{ are adjacent}\\
0, & \text{ otherwise.}
\end{array}\right.
\end{align*}
\item For any regular component $R_i$ and any 1-singular component $S_j$
\begin{align*}
e(R_i, S_j)=\left\{
\begin{array}{ccc}
1, & \text{ if } R_i,\, S_j \text{ are adjacent}\\
0, & \text{ otherwise.}
\end{array}\right.
\end{align*}
\end{enumerate}
\end{dfn}
\begin{figure}[t!]
\begin{center}
\includegraphics[width=.47\textwidth]{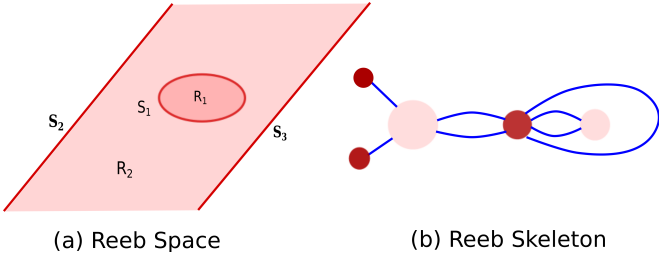}
\end{center}
\vspace*{-2ex}
\caption{(a) Reeb Space with self-adjacent 1-singular component (b)
  Corresponding Reeb Skeleton.}
\label{fig:reeb-skeleton2}
\end{figure}

\noindent
The regular and 1-singular components of
the Reeb Space are represented as the \emph{regular} and \emph{singular nodes},
respectively, in the Reeb Skeleton. \figref{reeb-skeleton} shows some
examples of Reeb Skeletons corresponding to the Reeb Spaces in \figref{jacobi-reeb}.
\figref{reeb-skeleton2} illustrates an example of the Reeb Skeleton
with a self-adjacent singular node.
Note that although the Reeb Skeleton gives a simple abstraction  of 0-connectivity in
the Reeb Space, it loses information of higher-dimensional
connectivities, like higher dimensional holes (tunnels, voids) in the
Reeb Space. But on the other hand, the Reeb Skeleton is extremely useful for
extracting any ``fork''-like structure (corresponding to a merge-split
feature)  in the Reeb Space. And we will see later
by a little simplification we can extract the most prominent merge-split
feature in the Reeb Skeleton and so in the Reeb Space.
Therefore, next we study properties of the Reeb Skeleton to simplify it
further.

\subsection{Simple Domains}
\label{sec:SimpleDomains}
We know from scalar fields that topologically simple domains have a useful property: the Reeb Graph is guaranteed 
to be a tree - i.e. the contour tree.  This not only enables more efficient computation, but also provides
straightforward mechanisms for feature extraction, simplification and
visualisation. Ideally, in multi-fields, the Reeb Space
would also be contractible to a point. But we show this is not true, in general.

\begin{figure}[t!]
\begin{center}
\includegraphics[width=8cm]{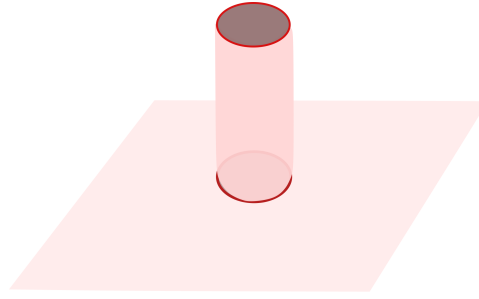}
\caption{Example of Reeb Space with a tunnel.}
\label{fig:reeb-tunnel}
\end{center}
\end{figure}

In topology, simple domains are characterised by
\emph{simply-connected space}. A topological space is simply-connected
if it is path-connected and every \textit{loop} in that space can be
continuously shrunk to a point without leaving the space.
In terms of homotopy theory this means a simply-connected space is
without any ``handle-shaped hole'' (as in \figref{reeb-tunnel}) or it has trivial fundamental group.  
For example, a sphere (that has a hollow center) is a simply-connected space
whereas a torus (that has a handle-shaped hole) is not. Even a simpler
topological space is known as \emph{contractible space} which
is homotopically equivalent to a point.  Note that a contractible space is
simply-connected, but the converse is not true. For example, a sphere
is simply-connected as every loop on it can be contracted to a point on
it, although the sphere is not a contractible space because of the center
hole in it. In the following lemma, we prove that the Reeb
Space corresponding to a map defined on a simply-connected domain is
simply-connected, but later we show it may not be contractible.

\begin{lemma}[\textbf{Simply-Connected}]
\label{lem:contractible}
The Reeb Space of a generic continuous map $f: \mathbb{X} \subseteq\mathbb{R}^3\rightarrow \mathbb{R}^2$ is simply-connected.
\end{lemma}
\noindent
\begin{proof} We consider any loop 
in the Reeb space $\RS$. Then, it lifts to an arc
in $\mathbb{X}$. But, every fiber of $q_f$ is connected,
and therefore, it lifts to a loop.
As $\mathbb{X}$ is simply-connected, this lifted loop
is null-homotopic. Therefore, its $q_f$-image
is also null-homotopic from the continuity of $q_f$. This means that $\RS$
is simply-connected. 
\end{proof}

\noindent
Therefore, if $f$ is good enough (for example,
triangulable or piecewise linear), then the Reeb space
is simply-connected. This implies that the 1st homology of the Reeb
Space also vanishes (or is the trivial group),
and therefore the Reeb space does not have a tunnel or $1$-dimensional
hole (i.e., a hole inside a circle $S^1$, e.g. \figref{reeb-tunnel}). Thus we
have the following theorem.
\begin{theorem}
\label{thm:tunnel}
The Reeb Space of a generic map $f: \mathbb{X}
\subseteq\mathbb{R}^3\rightarrow \mathbb{R}^2$ does not contain any
tunnel or $1$-dimensional hole.
\end{theorem}

\begin{figure}[t!]
\begin{center}
\includegraphics[width=6cm]{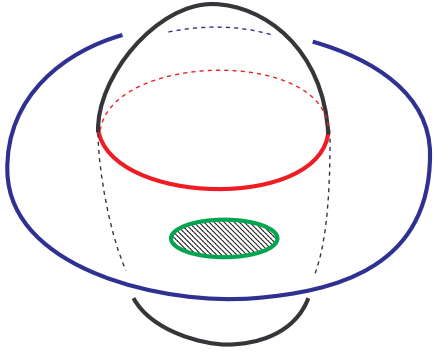}
\caption{Reeb Space with a void.}
\label{fig:hopf}
\end{center}
\end{figure}

On the other hand, for void or $2$-dimensional hole (i.e., hole inside a sphere $S^2$),
this is no longer true. We can construct a (piecewise
linear) map $f : \X \rightarrow \R^2$ whose Reeb space does have
a $2$-dimensional hole. For example, consider the Hopf
fibration $\Sp^3~\rightarrow~\Sp^2$ and its composition with
a standard projection $\Sp^2~\rightarrow~\R^2$. The resulting map
$\Sp^3~\rightarrow~\R^2$ is not generic, but perturbing
it slightly along its Jacobi set, we can obtain
a generic map $\Sp^3~\rightarrow~\R^2$, whose Reeb space is
the union of a $2$-sphere and an annulus attached
along the equator (and one boundary component of
the annulus). Then, by extracting a 3-ball
in the preimage of a two disk in the interior
of the annulus part, we get the desired map $\X~\rightarrow~\R^2$.
The Reeb space is the same space; the union of
$\Sp^2$ and an annulus (Figure~\ref{fig:hopf}).
Over each blue point lies a point (definite fold)
and it corresponds to a birth-death.
Over each red point lies a fiber as in Figure~\ref{fig:jacobi}(c) (with an
indefinite fold) and the splitting of a circle fiber
occurs. Over each green point lies a circle touching the
boundary of the domain cube $\X$. Thus,
over each point in the shaded disk bounded
by the green circle lies an interval.
 Note this disk is a subset of the
annulus part. Therefore, a Reeb Space of a multi-field
on a contractible domain may not be contractible and 
simplification of such space may not be simple as in 
the scalar case. 

According to Theorem~\ref{thm:tunnel},
we can conclude that each regular component of $\RS$ is
planar; i.e., each regular component is a disk possibly with
holes. For example, torus with holes (or a 1-dimensional hole as in Figure~\ref{fig:reeb-tunnel}) never appears!
This is essential in applying our simplification
rules for the Reeb Skeleton as will be discussed in
Section~\ref{sec:rules} (Figure~\ref{fig:rules-simplification}).
Next we focus on finding a criterion for detachability of such regular
components from the Reeb Space for simplifying the 
corresponding the multi-field. 

\subsection{Detachability}
\label{subsec:detach}
In the case of scalar field in a simply-connected domain,
the Reeb Space (Graph) is a contour tree and there always exists a leaf edge that can be
detached in a mathematically correct way,
unless the contour tree consists only of one edge.
We find similar criteria for defining
\emph{detachable} regular components in the Reeb Space.

We say that it is possible to detach a regular component from a Reeb Space
to obtain a simplified Reeb Space
if the multi-field corresponding to
the initial Reeb Space could be simplified 
to the multi-field corresponding  to the modified one, 
and then the regular component is said to be detachable from the Reeb Space.
Mathematically, any map could be simplified to a simpler map
in the following sense. Since $\R^2$ is contractible,
any two stable maps $f_0$ and $f_1 : \X \rightarrow \R^2$ are homotopic.
So, using singularity theory, we can show that
$f_0$ and $f_1$ are connected by a generic 1-parameter
family of maps. So, if we take an arbitrary stable
map as $f_0$ and a very simple map as $f_1$, then $f_0$
is simplified to $f_1$ after the generic 1-parameter
family. Such a 1-parameter family passes through
finitely many bifurcation parameters, and such
bifurcations can be classified \cite{Mata-Lorenzo-1989}.

\begin{figure}[h!]
\begin{center}
\includegraphics[width=0.47\textwidth]{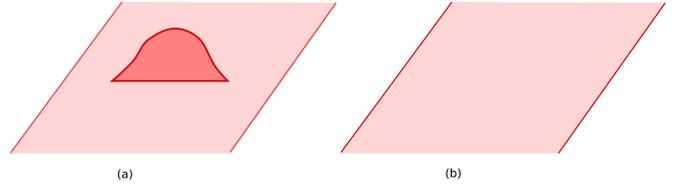}
\caption{Lip simplification: (a) Reeb Space with a lip, (b) Simplified
  Reeb Space.}
\label{fig:lips-dfn}
\end{center}
\end{figure}

Such transitions of the Reeb Spaces for generic smooth maps on a
closed 3-dimensional manifold into $\R^2$  have been studied in \cite{Mata-Lorenzo-1989},
although for maps on a 3-dimensional
manifold with boundary these results need further extension. In the current paper, we consider only   
a simple type of singularities and show that corresponding regular
component is detachable from the Reeb Space.
These components are known as \emph{lips}
and are defined as follows.
\begin{dfn}
\label{dfn:lips}
A lip is a regular component that is attached to the other
sheets of the Reeb Space exactly along one edge or an arc 1-singular component,
and it should not contain any vertex on the boundary,
except for the two cuspidal points (Figure~\ref{fig:lips-dfn}(a)).
\end{dfn}

\noindent
Next we prove the following lemma to show that the underlying map
corresponding to a lip can be simplified.
\begin{lemma}
\label{lem:detachability}
For a generic bivariate field $f: \X\subseteq \R^3\rightarrow\R^2$, a  ``lip'' can always be detached safely.
\end{lemma}
\noindent
\begin{proof}
If we have a lip in the Reeb Space, there are
three possibilities as shown in Figure~\ref{fig:lips}.
That is, we may consider the map near the inverse image of
the lip as a 1-parameter family of functions on a piece of surface.
Let $S_0$ be a ``piece of surface'' (a cylinder or a square as in Figure~\ref{fig:lips}),
and $f_t : S_0 \rightarrow \R,\, t \in \I$, 
be the 1-parameter family of height functions as in Figure~\ref{fig:lips}. 
Then, the original map $f$ is equivalent to the map $(x, t) \rightarrow
(f_t(x), t)$, $x \in S_0,\, t \in \I$, around the inverse image of a neighborhood of the lip by $q_f$. 
The Figure~\ref{fig:lips} presents the three such families of functions.
As we can see easily, these can be eliminated continuously,
by just shrinking the ``time interval'' for which a pair
of critical points appear.
\begin{figure}[h!]
\begin{center}
\includegraphics[width=8.3cm]{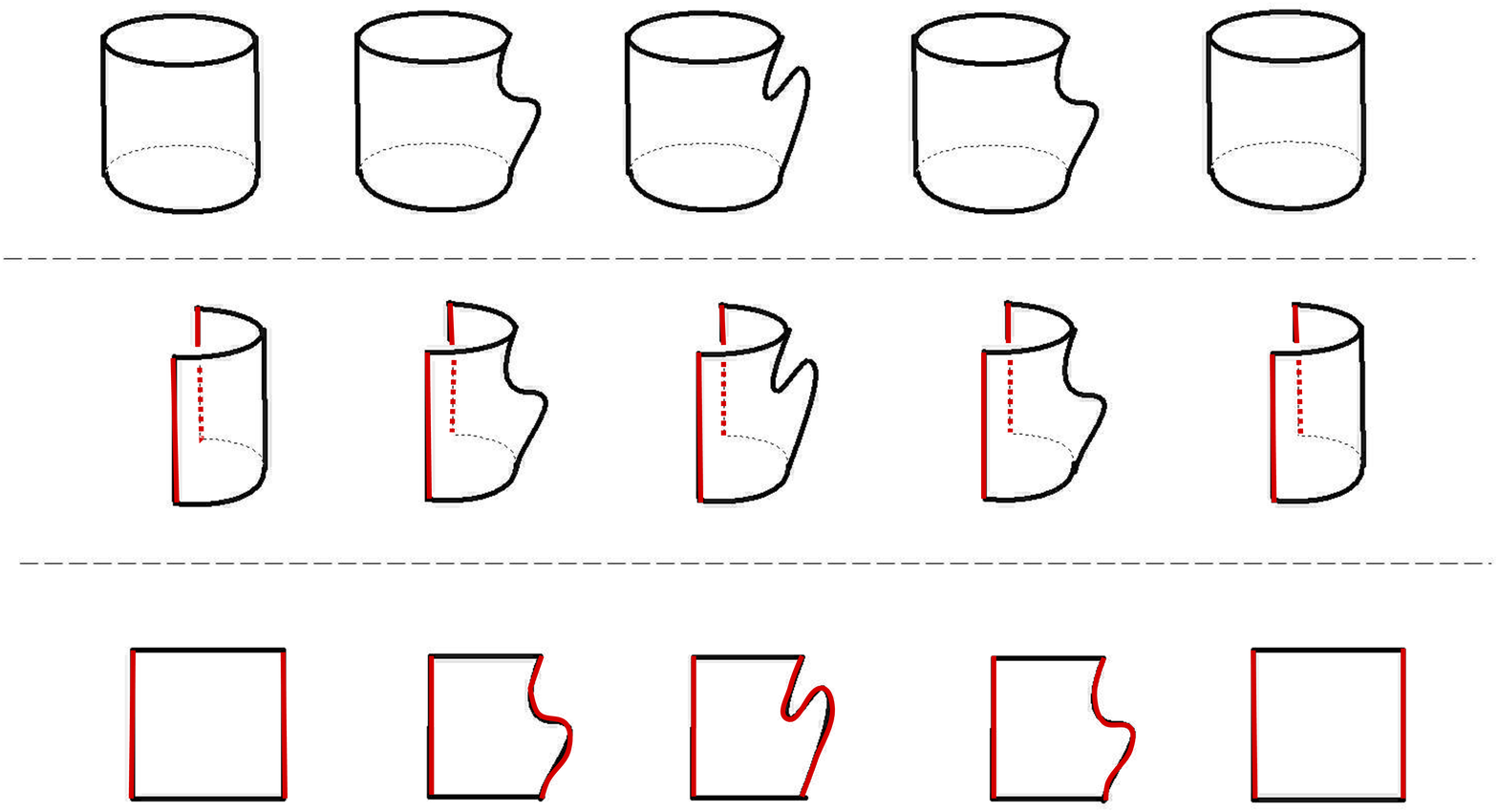}
\caption{The behavior of the stable map
  near the pre-image of a neighborhood of the lip. The red lines indicate the boundary of the domain.}
\label{fig:lips}
\end{center}
\end{figure} 
\end{proof}

\noindent
Thus, we see that lips are detachable and they can be simplified as in
Figure~\ref{fig:lips-dfn}. Therefore, we get our simplification rule for
detaching the lip components
as follows.\\

\noindent
\textbf{Simplification Rule:}
\emph{
Let $R_i$ be a detachable lip component of the Reeb
Space $\RS$. Then we simplify the Reeb Space by (i) deleting $R_i$ with
its adjacent boundary 1-singular component and
(ii) converting the attached arc 1-singular component (merge-split)
and two 0-singular components (cusp vertices) as regular.}

Next we discuss the Reeb Space (Skeleton) simplification based on the
rule developed in this section.
\section{Reeb Space Simplification and Measures}
\label{sec:Simplification}
In the real multi-field data because of noise 
very often there are ``lip''-like components which occlude the
original feature captured by the Reeb Space. 
Therefore it is important to simplify such components to
understand the topology of the underlying data.
Given that it is possible to detach such ``lip''-like regular components from the Reeb Space, we follow a similar strategy
to that used for the contour tree~\cite{CSv10}. There, a leaf edge was chosen for pruning
and removed from the tree. If as a result a saddle point became regular (i.e. 1-manifold), it too
was removed, simplifying the graph further. By tracking which leaves, saddles and edges are 
removed, the branch decomposition~\cite{PCS04} then gave a natural simplification 
hierarchy for any ordering of leaves.

In any Reeb space where \lemref{detachability} applies, we can use the same strategy, building
a simplification hierarchy in the process. \textcolor{black}{To do so, we simply choose a detachable 
component and remove it from the Reeb Space as described in the
simplification rule of Section~\ref{subsec:detach}.} 
We illustrate this process in \figref{demo-simplification}, where we
progressively remove detachable regular 
components from the Reeb Space, reducing the Jacobi Structure accordingly as much as desired.  As 
in leaf-pruning of contour trees, ``lip''-simplification reduces the number of regular components in the Reeb Space by one each 
time, and also remove components of the Jacobi Structure, guaranteeing that the number of steps
required is linear in the number of regular components of the Reeb Space. Moreover, the editing operations to update
the Reeb Space, Jacobi Structure and Reeb Skeleton are constant at
every step, \textcolor{black}{making the
simplification effectively linear (in the number of regular components) once the order of reduction is known.}
Therefore, we study different measures to associate with the regular
components (nodes) of the Reeb Space (Skeleton).

\subsection{Range Measure}
\label{sec:MeasurePersistence}
In simplifying the contour tree, Reeb Graph and Morse-Smale Complex, simplification can be defined 
by cancelling pairs of critical points according to an ordering given by a \emph{filtration} - i.e.
a sequence by which simplices are added to a complex.  For any given filtration, a unique ordering
exists, and the persistence of a feature is defined by the distance in the filtration between the
critical points defining the feature.  

For scalar data, however, the order in the filtration is dictated by the isovalues associated with
each vertex of the simplex, with the result that persistence can also be formalised as the isovalue
difference between the critical points that cancel each other.  In multi-fields, the persistence of 
a feature gives rise to tuples rather than a single value~\cite{Carlsson-MultiVariatePersistence},
which does not naturally give rise to a total ordering of the features.

This is however, not the only way to define a simplification ordering.
Carr et al. \cite{CSv10} 
showed that pruning leaves individually could be ordered by geometric properties such as area, 
volume etc. of the features defined by the contour tree. In this model, persistence is the vertical
height of a feature corresponding to a branch of the contour tree, and removing leaves can be done
with simple queue-based processing.
Recently, Duffy et al. \cite{DCM13} demonstrated that many properties of isosurfaces in scalar 
and multi-fields relate to geometric measure theory. In this model, statistical and geometric 
properties of a function are measured by integration over the range.
Following a similar approach we introduce a \emph{range measure} for
computing area of the regular components using the induced measure
from the range to the  Reeb Space. Note that, in general a regular component of a Reeb Space
is projected to the range with multiplicities: i.e., this map is an
immersion, but may not be injective.

Consider for example the Reeb spaces shown in \figref{jacobi-reeb} for bivariate volumetric 
maps. 
Mathematically, range measure of a regular component in the Reeb space $\RS$ is
defined as the area of the 2-dimensional sheets with respect to the measure induced from the usual
area measure of the range Euclidean space. 
The range measure of each regular component in the Reeb space is a fixed scalar value. Thus,
there is a unique induced ordering for simplification. If two 
components have identical range measure, some form of perturbation will be required to guarantee a 
strict ordering. 



\subsection{Geometric Measures}
\label{sec:geomMeasures}
Similarly, it is also possible to compute geometric properties of the regular components, either
in the domain, in the range, or in some combination of the two, using
geometric measure theory. 
As with the contour tree~\cite{CSv10}, obvious properties of interest include the measure of the 
region's boundary in the domain (contour length in 2D, isosurface surface area in 3D), the measure
of the region in the domain (area in 2D, volume in 3D), the measure of the function over the region 
(a generalisation of the volume in 2D, hypervolume in 3D), and so forth.  \textcolor{black}{However, as in that work, 
rules will be needed in each case for combination of measure with parents in the simplification 
hierarchy based on the theory in Section~\ref{subsec:detach}.}  

\subsection{Summary of Theoretical Contributions}
\label{sec:mathSummary}

We have now completed the theoretical groundwork for practical
simplification algorithm of Reeb Spaces. In particular our theoretical
results could be summarised as follows.

\begin{enumerate}\itemsep1pt
	\item	The Reeb Space consists of regular components corresponding to regions in the domain
			of the function, and singular components describing their relationships.
	\item	The Jacobi Set in the domain does not capture all of the structure of the singular
			components in the Reeb Space, and the Jacobi Structure is needed to do so.
	\item	The Jacobi Structure of the Reeb Space can be used to
          further collapse the Reeb Space into the Reeb 
			Skeleton.
	\item	Multifields with topologically simple domains can be
          simplified using a variation on the leaf-pruning used for
          contour trees. 
	\item	A Reeb Space measure and other geometric measures are introduced to guide
          the Reeb Space simplification process.
\end{enumerate}

We now turn to the practical and algorithmic part of this paper: how to simplify
the Joint Contour Net, an approximation of the Reeb Space. 

\begin{figure}
\begin{center}
\includegraphics[width=.35\textwidth]{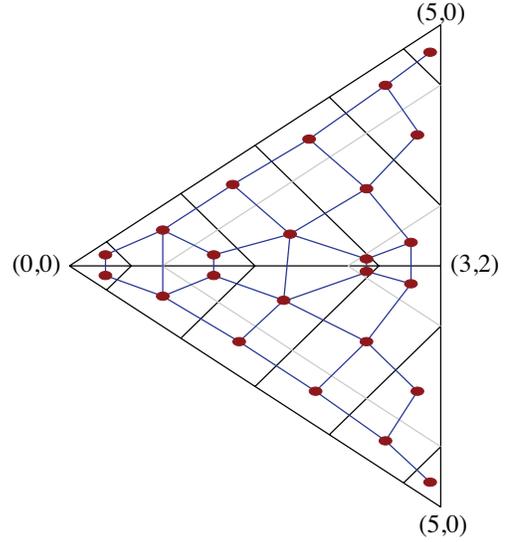}
\vspace{1cm}
\includegraphics[width=.35\textwidth]{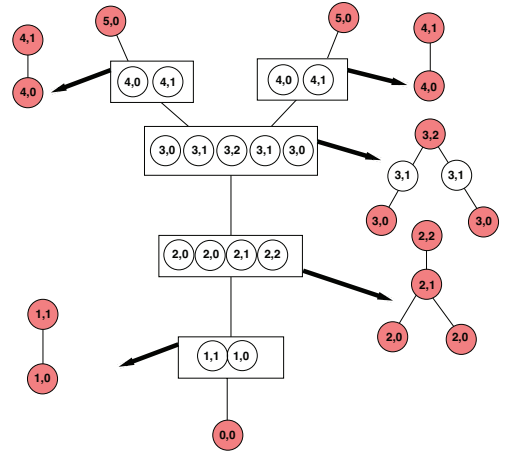}
\vspace{1cm}
\includegraphics[width=.35\textwidth]{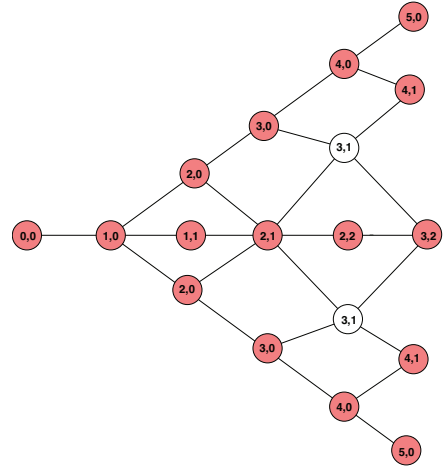}
\end{center}
\caption{ (top) The joint contour fragments and their adjacency graph
  for a PL-bivariate field defined by the values $\{(5,
  0), (0, 0), (5, 0), (3, 2)\}$ at the vertices of
  a mesh of two triangles. 
(middle) The Multi-Dimensional Reeb Graph constructed from the
  JCN. The critical nodes of the MDRG are the `red' nodes which form the
  Jacobi Structure.  
  (bottom)  Corresponding Joint Contour Net, with
  critical nodes from the MDRG marked in colour.
}
\label{fig:mdrg}
\end{figure} 

\section{Algorithm: Simplifying the Joint Contour Net}
\label{sec:SimplifyingJCN}
In this section, first we introduce the Joint Contour Net, a graph
data-structure that approximates the Reeb Space.
As described in \cite{2013-Carr-TVCG}, the Joint Contour Net is a
quantized approximation of the Reeb Space. Therefore, to avoid having
duplicate terminology we will use the same terminology for JCN
as what we have developed for the Reeb Space, namely, Jacobi Structure, Regular
Component, Singular Components, Reeb Skeleton etc.

\subsection{Joint Contour Net}
\label{sec:jcn}
The Joint Contour Net (JCN) \cite{2013-Carr-TVCG, 2012-Duke-VisWeek} approximates the 
Reeb Space $\RS$ of a multi-field $f=(f_1,\,f_2,\,\ldots,\,f_r):\X\subset\mathbb{R}^d\rightarrow \mathbb{R}^r$
in a $d$-dimensional interval  $\X$. Let
$\tilde{f}=(\tilde{f}_1,\,\tilde{f}_2,\,\ldots,\,\tilde{f}_r):
M\rightarrow \mathbb{R}^r$ be a piecewise-linear (PL) approximation of $f$ corresponding to a
mesh $M$ of $\X$. The idea of computing the JCN is based on
quantization of the fiber-components of $\tilde{f}$. The JCN $\tilde{f}$ with a
quantization level (or level of resolution) $m_Q$ is denoted as
$\JCN(\tilde{f}, m_Q)$, where $m_Q$ refers to how fine the rectangular mesh
for the range is.

A \textit{quantized level set} of $\tilde{f}_i$ at an isovalue $h \in
\mathbb{Z}/m_Q$ is denoted by $Q\tilde{f}_i^{-1}(h)$ and is
defined as:
$ Q\tilde{f}_i^{-1}(h):=\big\{x\in M : (\frac{1}{m_Q}) \operatorname{round}(m_Q\tilde{f}_i(x))=h\}$.
A connected component of the quantized level set in the mesh is called
a \textit{quantized contour} or a \textit{contour slab}. The part of the
contour slab in a single cell of the mesh is called a \textit{contour
  fragment}. 
  
Now the first step of the JCN algorithm constructs all the
contour fragments corresponding to a quantization of each component
field. In the second step, the \textit{joint contour fragments} are computed by
computing the intersections of these contour fragments for the component
fields in a cell. The third step is to construct an adjacency graph of these
joint contour fragments where a node in the graph corresponds to a joint
contour fragment and there is an edge between two nodes if the corresponding
joint contour fragments are adjacent. Finally, the JCN is obtained by
collapsing the neighbouring redundant nodes with identical isovalues.
Thus, each node in the JCN corresponds to a \textit{joint contour
  slab} (or \emph{quantized fiber-component}) and an edge represents the
adjacency between two quantized fiber-components (with quantization level $m_Q$) of $\tilde{f}$.

Note that one can build a multi-resolution JCN by increasing or
 decreasing the quantization level using a scaling factor for the
 ranges of the component fields. An example of a small JCN is given in \figref{mdrg}, but
 we refer the interested reader to \cite{2013-Carr-TVCG} for details.
 The following lemma shows that in the limiting case, when the
 quantization level increases and the domain-mesh becomes more
 refined, then the JCN converges to the corresponding Reeb Space.

\begin{lemma}[\textbf{Convergence}]
\label{lem:convergence}
Let $f : \X \subset \R^d \to \R^r$, $d \geq r$, be a tiangulable continuous 
multi-field with the Reeb space $\RS$. 
Choose an increasing sequence of quantization levels $\{m_Q^{(n)}\}$ for $f$ 
such that $m_Q^{(n)}$ is an integer multiple of $m_Q^{(n-1)}$ for each $n$ 
and $\displaystyle\lim_{n \to \infty}m_Q^{(n)} = \infty$. 
Furthermore, let $\{M_n\}$ be sequence of sufficiently fine meshes 
of $\X$ such that $M_n$ is a refinement of $M_{n-1}$ for each $n$ and 
$\displaystyle\lim_{n\to \infty} d(M_n) = 0$, where $d(M_n)$ stands for 
the maximum of the diameters of the cells of $M_n$. 
Finally, let $f^{(n)} : M_n \to \R^r$ be the PL map associated with $f$ 
corresponding to the mesh $M_n$. 
Then the sequence $\left\{\JCN(f^{(n)}, m_Q^{(n)})\right\}$ 
converges to $W_f$.
\end{lemma}
\noindent
\begin{proof} 
Hiratuka et al. \cite{Hiratuka2013} show
for a PL map $f : A \to B$ of a compact polyhedron $A$ into
another polyhedron $B$, if we subdivide the range polyhedron $B$
appropriately, then $A$ is subdivided accordingly and the quotient map
$q_f : A \to \RS$ to the Reeb Space $\RS$ is triangulable with respect
to the triangulations. In the proof, it is also shown that the inverse
image by $q_f$ of a small regular neighborhood of a vertex $v$ in
$\RS$ is always a regular neighborhood of $(q_f)^{-1}(v)$ in $A$. This
implies that if the quantization level is high enough, then the
quantized fiber-component is actually a regular neighborhood of the
central fiber-component. Consequently, we have a natural embedding
$\rho_0: JCN_0^{(n)} \to \RS$, where $JCN_0^{(n)}$ is the set of
vertices of the Joint Contour Net $\JCN(f^{(n)},m_Q^{(n)})$ for
sufficiently large $n$. (For each quantized
fiber-component, associate the central fiber-component.) Furthermore,
as is shown in \cite{Hiratuka2013}, this embedding preserves the
adjacencies. This implies that the embedding $\rho_0$ extends to
an embedding $\rho : \JCN(f^{(n)},m_Q^{(n)}) \to \RS$. Hence, the required result holds,
since the triangulations of $\JCN(f^{(n)},m_Q^{(n)})$ and $\RS$
becomes finer and finer as $n$ increases.

If $f$ itself is not a PL map, then we can consider its triangulation
$g : A \to B$ and obtain the required result for the triangulation. As
the domain $\X$ is compact, this implies the same consequence for the
original map $f$ as well. This completes the proof.
\end{proof}

Next we see that the simplification will have four stages: 1. extraction of the 
Jacobi Structure from the JCN, 2. computing regular and singular
components for construction of the Reeb Skeleton,
3. computation of measures for each regular node in the Reeb Skeleton,
and 4. simplification by pruning nodes corresponding to the ``lip'' components.  In practice, the first stage is the most difficult - identifying the regular 
components, and this requires an intermediate data-structure, which we introduce now.

\subsection{Multi-Dimensional Reeb Graphs}
\label{sec:MDRG}
The first step in detecting and analysing the Jacobi Structure is to identify the
nodes in the JCN that capture changes in the topology - i.e. the quantized 
representatives of the Jacobi Structure.  To do so, we exploit a simple property
of the JCN - that the slabs can be arranged hierarchically, with the levels of the
hierarchy corresponding to the individual fields. At the highest level of the 
hierarchy, the slabs are only defined by field $f_1$, and are therefore equivalent
to interval volumes: as such, we can compute the Reeb Graph for field
$f_1$ (see Figure \ref{fig:mdrg}~(middle)). 

\begin{algorithm}
\caption{{\sc CreateReebGraph}$(G, f_i)$}
\label{alg:rg}
{\bf Input:} A subgraph $G$ of $JCN$ and a chosen field $f_i$\\
{\bf Output:} The Reeb Graph $RG$ with respect to field $f_i$
\begin{algorithmic}[1]
\State Create Union-Find Structure $UF$ for field $f_i$.
\State For each adjacent $g_1,g_2 \in G$ with $f_i(g_1) = f_i(g_2)$, UFAdd($g_1,g_2)$
\For{ each component $C_l$ in UF}
\State Create a node $n_{C_l}$ in $RG$
\State Map graph node-id(s) and field-values from $G$ to $n_{C_l}$
\EndFor
\State Order nodes ${n_{C_1},\ldots,n_{C_n}}$ according to $f_i$ field values.
\For{edge $e_1e_2$ in $G$}
\If {$e_1, e_2 \in$ components $C_j \neq C_k$ and $f_i(e_1) \neq f_i(e_2)$}
\State Add edge $e(n_{C_j}, n_{C_k})$ in $RG$ if not already present
\EndIf
\EndFor
\State \Return{$RG$}
\end{algorithmic}
\end{algorithm}

Each slab (i.e. interval volume) of $f_1$ can be broken up into smaller slabs with respect to
field $f_2$ in a similar way (which form a subgraph $G$ in the JCN), and
the Reeb graph for these slabs computed similarly, as
shown in \algoref{rg}. Proceeding recursively, we then compute a hierarchy of Reeb graphs, 
each of which represents the internal topology of a slab of the parent Reeb Graph  
with respect to the child's field.  We call this hierarchy the \emph{Multi-Dimensional
Reeb Graph} or MDRG and denote this as $\MDRG$.

Computing the MDRG is straightforward once the full JCN has been extracted: we 
start with the JCN and compute the Reeb Graph for property $f_1$ by performing
union-find processing over the nodes of the JCN.  This breaks the JCN into subgraphs
corresponding to slabs in the Reeb Graph of property $f_1$. The MDRG for each 
subgraph is then computed recursively, and stored in the node of the parent Reeb 
Graph to which its slab corresponds. In the process, the slabs get separated out into 
smaller and smaller components.

\begin{algorithm}
\caption{{\sc MultiDimensionalReebGraph}$(G, f_i,\ldots, f_r)$}
\label{alg:mdrg}
{\bf Input:} Graph $G$, fields $f_i, \ldots, f_r$
{\bf Output:} MDRG $M$
\begin{algorithmic}[1]
\If {$i \leq r$}
\State	Let $R = $ CreateReebGraph($G, f_i$) 
\State Store $R$ as root node of $M$
\For {Each slab $s$ of $R$}
\State Extract subgraph $G_s$ of nodes of $G$ belonging to $s$ in $R$ 
\State Compute $M_s = $ \small{MultiDimensionalReebGraph($G_s, f_{i+1}, \allowbreak \ldots, f_r$)}
\State	Store $M_s$ at node $s$ of $R$
\EndFor
\State \Return $M$
\Else
\State \Return $M = \emptyset$
\EndIf
\end{algorithmic}
\end{algorithm}

We state this as an algorithm in \algoref{mdrg} and illustrate with a bivariate
field in \figref{mdrg}.   This algorithm is stated recursively for simplicity, but can also
be implemented with queue processing for speed. Moreover, the division of subgraphs
at each level into slabs can be performed more efficiently by exploiting the connectivity 
already encoded in the JCN.  

The principal value of the MDRG is that every node of the JCN in the Jacobi Structure is
guaranteed to be a critical node of the finest-resolution Reeb
Graphs (denoted as the critical nodes of the MDRG). This immediately gives a method of computing the Jacobi Structure once the MDRG
is known \cite{2014-EuroVis-short}.

\subsection{Jacobi Structure Extraction}
\label{sec:JacobiStructureExtraction}
Since every node belonging to the Jacobi Structure is guaranteed to appear as a critical node
of the lowest level of an MDRG, the initial stage in Jacobi Structure extraction is simply to
mark these nodes.  Unmarked nodes are then guaranteed to be regular, and can be collected
into regular components.  Once this has been done, any remaining nodes that are adjacent
to each other and to the same set of regular components are identified, as these form a 
 1-singular component between the regular components.

The first stage of this can be seen in \figref{mdrg}, where the critical nodes of the lowest
level of the MDRG together mark all of the Jacobi Structure nodes in the JCN (in colour).

\subsection{Reeb Skeleton Construction}
\label{sec:reeb-skel}
\begin{figure}[t!]
\begin{tabular}{|c|c|}
\hline
\subfloat[\label{fig:6a}]{\includegraphics[height=3.5cm]{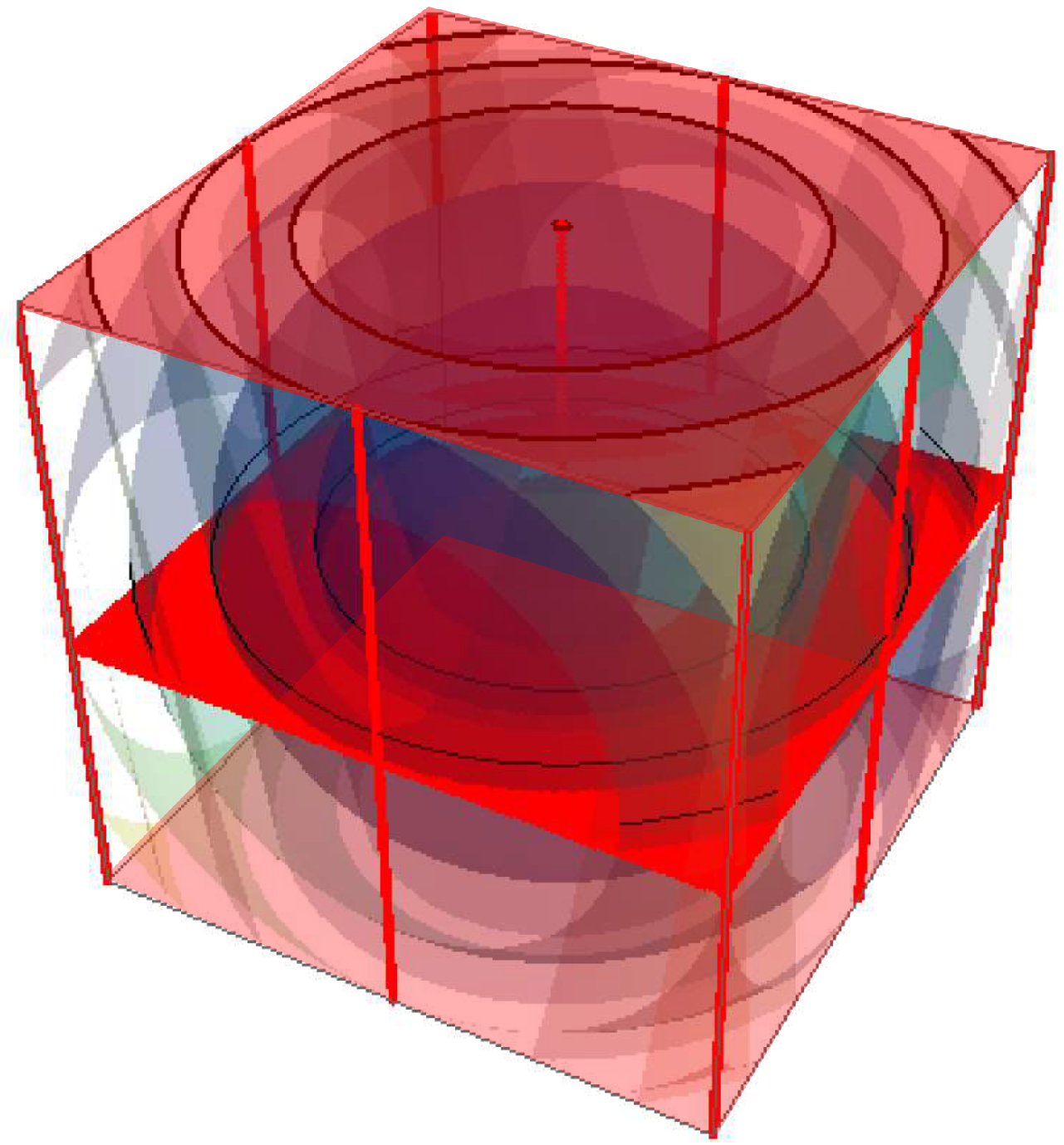}} & \subfloat[\label{fig:6b}]{\includegraphics[height=3.5cm]{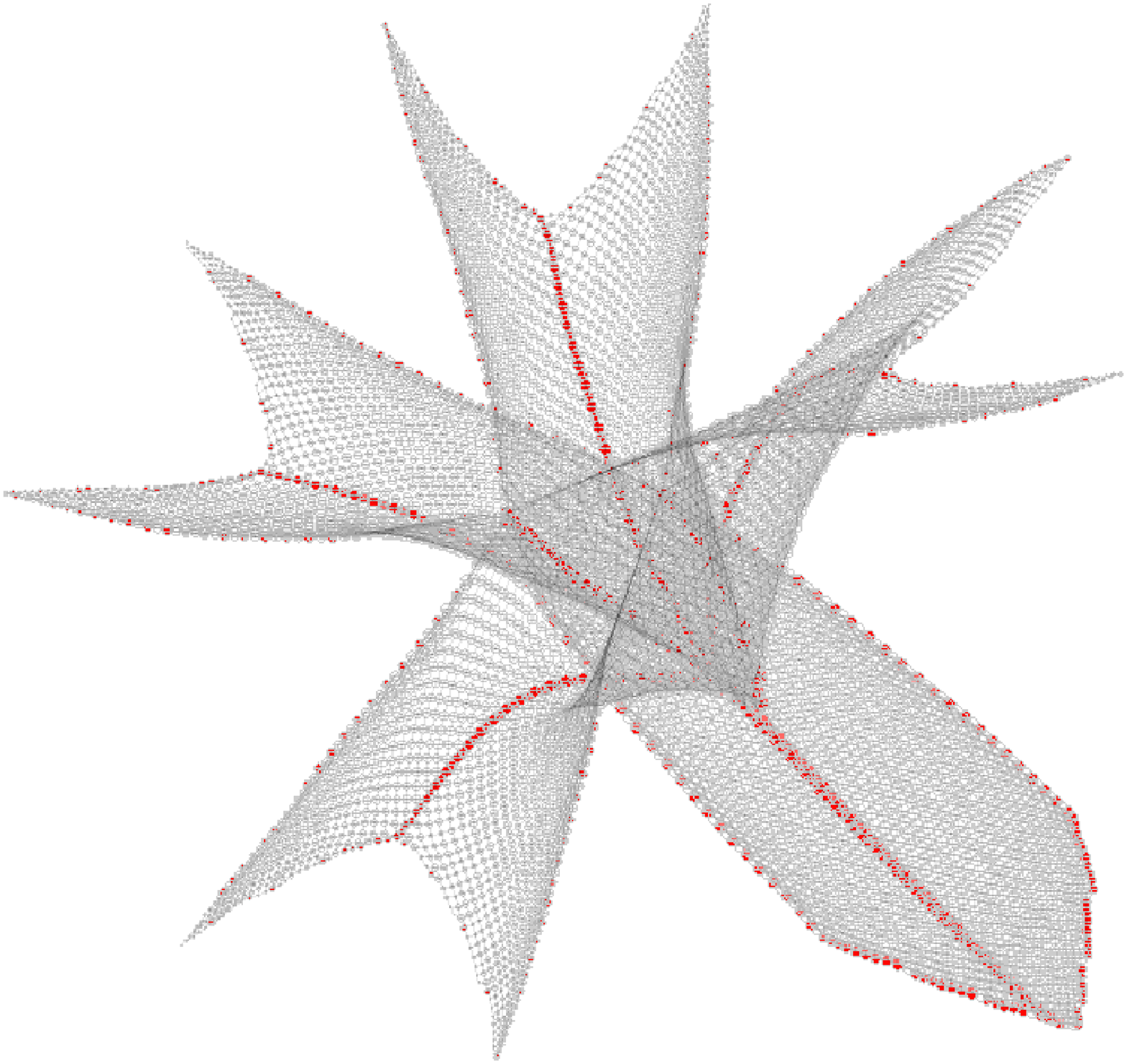}}\\
\hline 
\subfloat[\label{fig:6c}]{\includegraphics[height=3.5cm]{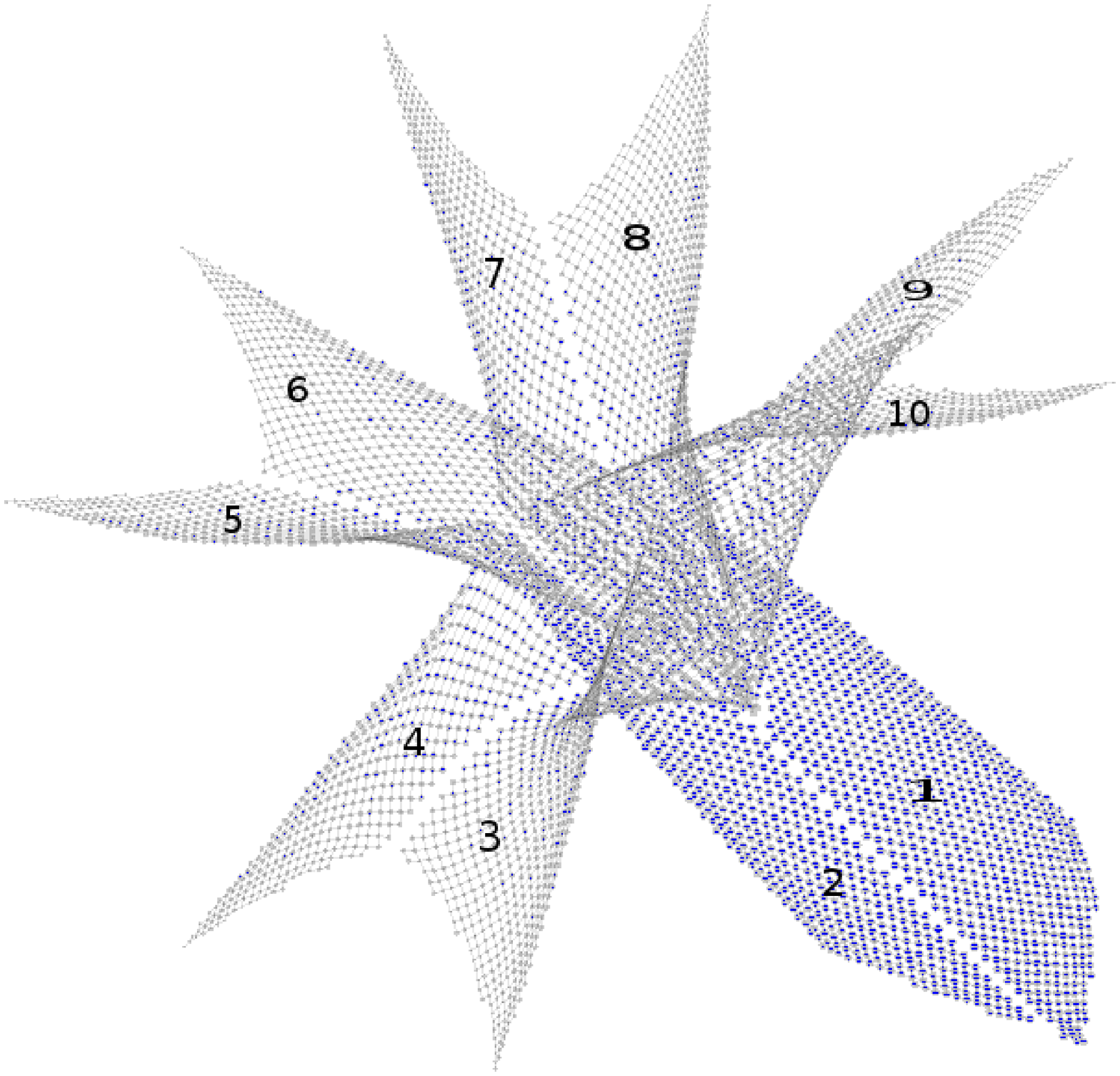}}& \subfloat[\label{fig:6d}]{\includegraphics[height=3.5cm]{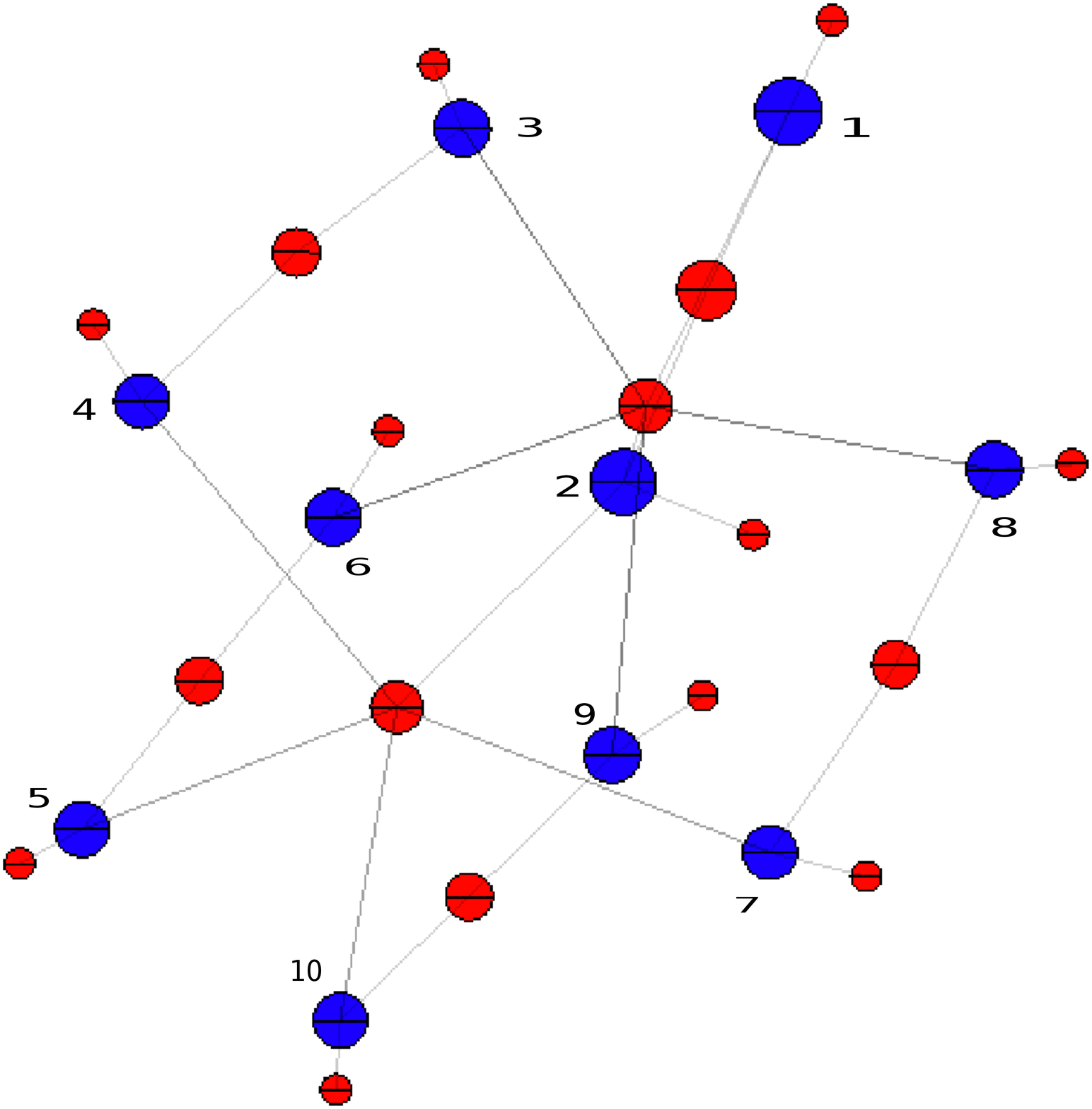}}\\
\hline
\end{tabular}
\caption{(a) Bivariate Field: $(x^2+y^2-z$, $x^2+y^2+z^2)$ in a box
  $[-5,\,5]^3$, the `red' components are the Jacobi Set, (b) the Joint
  Contour Net with the Jacobi Structure (in red), (c) Regular
  Components, (d) the Reeb Skeleton.}
\end{figure}
\label{fig:reeb-skeleton-demo}
Once we have extracted the Jacobi Structure for the JCN, it is straightforward to compute the
corresponding Reeb Skeleton by creating a single node for each regular or 1-singular component,
and connecting them using the adjacency of components in the Jacobi
Structure. In \figref{reeb-skeleton-demo} we see an example of the Reeb
Skeleton construction for a volumetric bivariate field. We note, in
this case, the Reeb Skeleton has no
detachable lip-like regular node whereas the Reeb Skeleton
\figref{demo-simplification} (d) has such detachable nodes. 

Now in the simplification algorithm, the order in which detachable
Reeb Skeleton nodes are removed  is determined by the metrics associated with those
nodes. Computation of such metrics are described next.

\begin{algorithm}
\caption{\sc{ SimplifyReebSpace}}
\label{alg:simplify-jcn}
{\textbf{Input:}} JCN $\JCN$\\
{\textbf{Output:}} Reeb Skeleton $\RK$
\begin{algorithmic}[1]
\State Build MDRG $\MDRG$ and Jacobi Structure $\JS$ from JCN $\JCN$.
\State Partition $\JCN$ into disjoint regular components $C = \{R_1, \ldots, R_m\}$ by deleting $\JS$ from $\JCN$. 
\State Partition $\JS$ into disjoint 1-singular components $\{S_1, \ldots, S_n\}$ based on adjacency to regular components in $C$.
\State Use adjacencies of $\{R_1, \ldots, R_m, S_1, \ldots, S_n\}$ to
construct $\RK$ (following defintion~\ref{dfn:reeb-skeleton}). 
\State Push ``detachable'' nodes of $\RK$ on priority queue $PQ$ with priority determined by geometric measures.
\While {$PQ$ not empty (or priority is below a threshold value)}
\State Pop node $r$ from queue
\State Prune $r$ from $\RK$ 
\EndWhile
\State \Return {Simplified Reeb Skeleton $\RK$.}
\end{algorithmic}
\end{algorithm}

\subsection{Computing Simplification Metrics}
\label{sec:metric}
Our simplification algorithm 
can use any desired measure of importance for components of the Reeb
space, including but not limited to:

\begin{itemize}
\item \textbf{Range measure.} As described in \subsecref{MeasurePersistence}, we can measure
the size of the regular components by the induced measure of the
range. This is easy to approximate - in this case, by the number 
of unique JCN slabs (i.e. pixels in the range) that map to a given
regular component.

\item \textbf{Surface area.} A regular component of the Reeb space is separated from
other regular components by one or more singular components in the Jacobi Structure. Since
the regular components correspond to features and the singular components to boundaries
between features, we can associate the area of the bounding surface with the regular component
for the purpose of simplification.  For the JCN, we can approximate this with the
surface area of the fragments adjacent to the bounding region.

\item \textbf{Volume.} Similarly, we can measure volume in the domain for each feature represented
by a regular component, and approximate it by summing the volume of fragments mapping 
to a given regular component.

\item \textbf{Other measures.}  As shown by Duffy~\etal~\cite{DCM13} and Carr~\etal~\cite{CSv10},
almost any geometric or other property of features can be used for simplification provided that it is
correctly approximated and suitable rules for composition during 
simplification are established.
\end{itemize}

\subsection{Simplifying the Reeb Skeleton}
\label{sec:rules}
\begin{figure}[h!]
\begin{center}
\includegraphics[width=5.3cm]{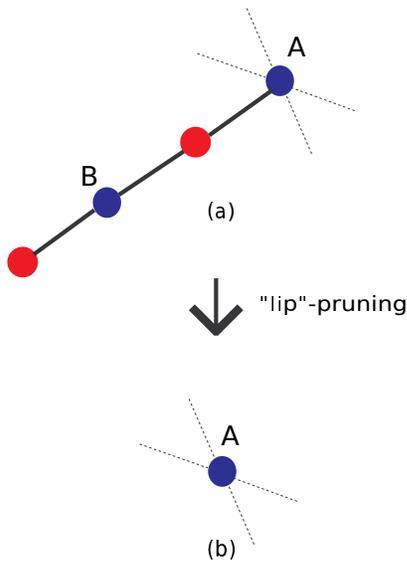}
\caption{Simplification rule for detaching a lip-like node (regular
  node B in (a)) from the Reeb skeleton.}
\label{fig:rules-simplification}
\end{center}
\end{figure}

The lip-simplification rule of the Reeb Space, as described in
Section~\ref{subsec:detach}, can be translated similarly in the
corresponding Reeb Skeleton. We note, according to Theorem 4.4,
 each regular component of our Reeb Space
 is always a disk possibly with holes. Therefore, the degree 2 regular
 node B as in a Reeb Skeleton \figref{rules-simplification}(a) is always a
 lip-like detachable node. \figref{rules-simplification}(b) shows the simplified Reeb
 Skeleton after pruning the lip-like node and attached singular nodes.

Finally, we give simplification strategies of the
Reeb Skeleton, based on geometric and range measures of the
components. The Reeb skeleton simplification method simplifies the Reeb skeleton
and the corresponding Reeb space given a threshold (between 0 and 1) adapting the previous approaches
in the literature \cite{2004-Carr-simplification, 2012-Tierny-tvcg}. The threshold represents a
``scale'', under which detachable regular nodes of the Reeb skeleton
are considered as unimportant (noise).
The threshold is expressed as a fraction of the range of the metric used. It can vary from 0 (no simplification) to 1
(maximal simplification).

\begin{figure}[t]
\begin{center}
\includegraphics[width=.47\textwidth]{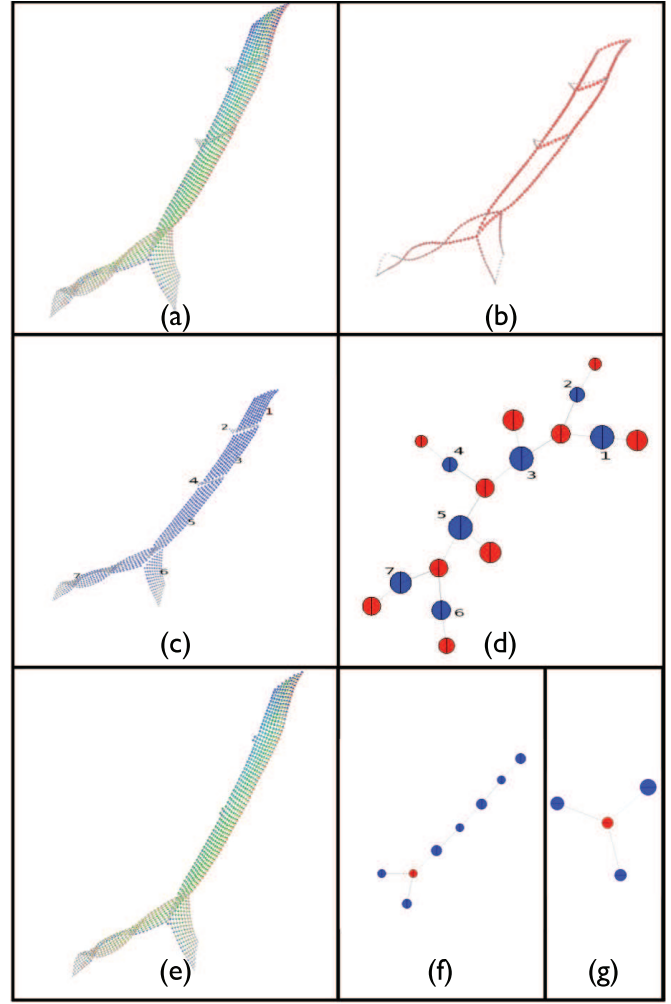}
\caption{(\textbf{Simplification Demo}) (a) Original  JCN/Reeb Space of bivariate field (Paraboloid, Height) 
  (b) Jacobi Structure, (c) Regular components (d) Reeb Skeleton (`blue'
  corresponds to regular components and `red'
  corresponds to adjacent 1-singular components)  (e)
  Simplified JCN
  (f)-(g) Simplified Reeb Skeleton using range measure.}
\label{fig:demo-simplification}
\end{center}
\end{figure}

\figref{demo-simplification} demonstrates the simplification
of components from the  Reeb Space, of  an unstable bivariate
volumetric data. We use range measure for ordering the components.
We note, regular nodes 4  and 2 in \figref{demo-simplification} are not strictly the lips according to our
definition of lips. However, a perturbation can be applied first 
to convert such components to lips and then lip
simplifications can be applied. In \figref{demo-simplification} we apply our lip-simplification rule  
directly at the regular nodes 4  and 2, sequentially (node with
smaller measure is pruned first).

\section{Implementation and Application}
\label{sec:Implementation}

We implement our Reeb Space or JCN simplification algorithm using the Visualization Toolkit (VTK)
\cite{vtk}. Details of the JCN implementation can be found in \cite{2013-Carr-TVCG,
  2012-Duke-VisWeek}. Our Reeb Space simplification implementation takes the JCN (a vtkGraph
structure) as input and builds four filters: (1) The first filter computes the
Jacobi Structure by implementing the Multi-dimensional Reeb graph
algorithm, (2) The second filter builds the Reeb Skeleton structure by
partitioning the JCN, (3) The third filter implements persistence and geometric measures and (4) The fourth filter implements the
simplification rules of the Reeb Skeleton.
We use a vtkTree structure to store the Multi-Dimensional Reeb Graph (MDRG): 
Reeb graphs at each level of MDRG are stored in a vtkReebGraph structure. For capturing 
the Reeb skeleton we use the vtkGraph structure. 

\begin{table}[h!]
 \begin{centering}
 \caption{Data Statistics}
\scalebox{0.65}{
\begin{tabular}{lccccccc}
\hline
datasets        & spatial-dimensions & slab widths & no. of nodes (JCN) & no. of edges (JCN)\\ 
\hline 
(Circle, Line) &(29, 29, 1)	           &(1, 1)	           &500		&1057	\\
(Paraboloid, Height)	&(40, 40, 40)		           &(1, 1)		&1260		&2383		\\
(Sphere, Height)	&(40, 40, 40)		           &(1, 1)		&1308		&2428		\\
(Paraboloid, Sphere)	&(40, 40, 40)		           &(1, 1)	&6554		&12795		\\
(Cubic, Height)	&(40, 40, 40)		           &(1, 1)	&3149		&5928		\\
\end{tabular}
}
\label{tab:perf-mdrg}
\end{centering}
\end{table}
A force-directed graph-layout from the OGDF - an Open Graph Drawing
Framework \cite{ogdf} strategy has been used for the graph
visualization as shown in the demonstrations and outputs.
We run our implementation on different synthetic and simulated data sets for
testing the performance.  In \tabref{perf-mdrg} the synthetic data sets are 
labelled by the combination of scalar fields used: Circle: $x^2+y^2$, Line: $y$, 
Sphere: $x^2+y^2+z^2$, Paraboloid: $x^2+y^2-z$, Height: $z$ and Cubic: $(y^3-xy+z^2, \,
  x)$. Circle and Line are in the 2D-box $[-5,\, 5]^2$ and other fields are
  considered in the 3D-box $[-5,\, 5]^3$. 

\begin{table}[h!]
 \begin{centering}
 \caption{Performance results for Simplification}
\scalebox{0.7}{
\begin{tabular}{lccccccc}
\hline
Data                         & Spatial & Slab  &Jacobi &  Reeb &  \\ 
                         & Dimensions  & Widths  &Structure &  Skeleton & Simplification \\ 
\hline 
(Circle, Line)	                &(29, 29, 1)	     &(1, 1) &0.06s & 0.242s	&0.00s\\
(Paraboloid, Height)		&(40, 40, 40)	     &(1, 1) &0.10s	   &1.97s	&0.00s\\
(Paraboloid, Sphere)		&(40, 40, 40)	     &(1, 1) &0.80s	   &33.02s	&0.00s\\
Nucleon &(40, 40, 66) &(8,2) & 0.45s& 48.91s & 0.45s\\
\end{tabular}
}
\label{tab:perf-simpl}
\end{centering}
\end{table}
\paragraph{Performance results} Table~\ref{tab:perf-simpl} shows the performance
results of the JCN and MDRG algorithms for some simulated
data. All timings were performed on a 3.06 GHz 6-Core Intel Xeon with 64GB memory, running
OSX 10.8.5, and using VTK 5.10.1. 

The number of nodes in the MDRG is actually the number of Reeb graphs
computed by the MDRG Algorithm~\ref{alg:mdrg}. From the table it is clear that
performance of the MDRG algorithm is quite impressive for these
simulated data. The complexity of the CreateReebGraph on a graph with $n$ nodes is $O(n+p\log n)$ 
which is the complexity of a sequence of $p$ UF operations (here,
$p\leq n$) \cite{1975-tarjan}.

\subsection*{Nuclear Scission Data}
\begin{figure*}
	\centering
	 \includegraphics[width= 0.71\textwidth]{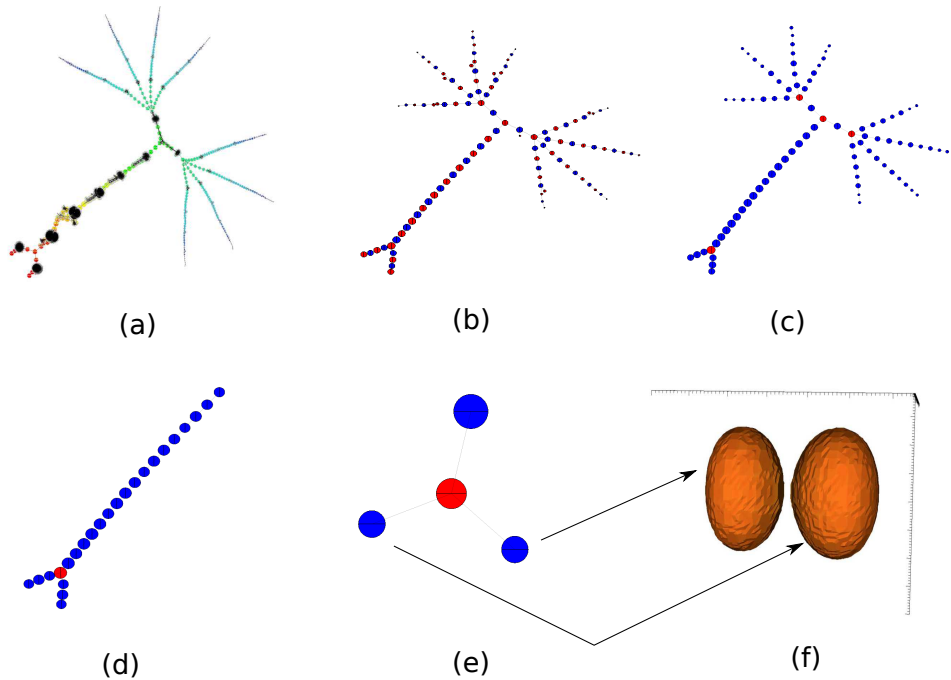}  \\
	\caption{ This figure shows the simplification process of the
          JCN corresponding to a ``nuclear scission'' data set used in \cite{2012-Duke-VisWeek}. 
	The process includes: (a) The original JCN graph; (b) The Reeb
        Skeleton; (c) A modified representation of the Reeb Skeleton
        by changing the colour of the degree 2 singular nodes as
        ``blue'', since they capture  only minor topological information, (d)
        Simplified Reeb Skeleton after applying lip-pruning using
        the range measure for ordering the nodes (e) Resultant Reeb
        Skeleton is represented  as the ``Y''-fork; 
	(f) Geometry corresponding to the ``scission'' point.   }
	\label{fig:nuclear}
\end{figure*}
In a previous application paper~\cite{2012-Duke-VisWeek}, the JCN was
applied to nuclear data set (time-varying bivariate field of proton and neutron densities) and used to visualise the scission points in
high-dimensional parameter spaces. Here the scission refers to the
point where a single plutonium nucleus breaks into two fragments.
However, this was based on visual analysis, and was complicated by a 
number of artefacts such as the recurring chains of star-like ‘motifs’ within 
the JCN. Moreover, the eight corners of the domain boundary induced 
eight corresponding small sets of features in the JCN. 

Here we apply our simplification algorithm to
overcome these artefacts and preserve the principal topological
feature. Note that apparently there are two red nodes that are adjacent to
5 blue nodes in the Reeb Skeleton \figref{nuclear}(b). This never
happens if the given multi-field is stable, so this is an example of
an unstable bivariate field in 3-dimensional interval. We demonstrate the process of simplification for one of these 
scission data sets in \figref{nuclear}. Note that the final simplification
results a simple Y-fork where two regular nodes correspond to the separated
nuclei, while the third represents the exterior.

\section{Conclusions and Discussions}
\label{sec:Conclusions}
In this paper, we provide a rigorous mathematical and computational foundation of
multivariate simplification based on lip-pruning from its Reeb Space 
and this generalises approaches that are effective for scalar fields.
We note, lip-simplification can be applied only when there is a
lip-component in the Reeb Space and might not always be possible, but
nevertheless, it is quite effective when applied to real data sets,
which usually contain a lof of noise (as demonstrated in
\figref{nuclear}). The Jacobi Structure
that characterises the Reeb Space is richer than the Jacobi
Set and decomposes the Reeb Space. This is proved to be a useful
property for the simplification procedure.
In addition, we have shown how to extract a reliable
approximate Jacobi Structure and Reeb Skeleton from the JCN that can be simplified to 
improve the use of the JCN for multivariate analysis, and illustrated this with analytical
datasets and a real-world data. However, there are few open issues which
need to be addressed in future research.

\begin{itemize}\itemsep1pt
\item \textbf{False lips:} Currently using our lip simplification
  approach we are not able to distinguish or simplify the false-lips,
  similar as in \figref{falselip}. This is because our Reeb Skeleton cannot compute
the multiplicity of the adjacency between a regular and a 1-singular
node which will be important for detecting such a false lip in the Reeb Space. So, in our current simplification we assume no false lip 
appears and we can simplify only the ``genuine'' lips as defined in Definition~\ref{dfn:lips}.
\begin{figure}[h!]
	\centering
	 \includegraphics[width= 0.22\textwidth]{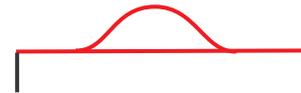}  \\
	\caption{False lip}
	\label{fig:falselip}
\end{figure}

\item \textbf{Discontinuity in components of Jacobi Structure:}
  Computed 1-singular components of
  the Jacobi Structure in the JCN may be discontinuous because of the degeneracy, as all degenerate singular points may not be captured by the critical nodes of
  MDRG. Moreover, a choice of the quantization level may also
  result in the discontinuous Jacobi Structure in the JCN. 

\item \textbf{Further Structures in the Reeb Skeleton:} In the current
  implementation of the Reeb Skeleton we have considered only the
  regular and the 1-singular components and their adjacency graph. But, there
  is further hierarchy possible in
  the Jacobi Structure. For more than bivariate case,
  singular components can be decomposed into lower
  dimensional manifolds (strata) and can be represented in the Reeb Skeleton,
  hierarchically. However, detecting such lower dimensional strata needs
  further theoretical analysis and an algorithm for detecting them.
\end{itemize}

\noindent
Apart from these issues, in the future, we intend to work on further simplification and acceleration of these techniques, 
and on alternate methods for Reeb Space computation and / or approximation.  We also expect
to examine more data sets from multiple domains, now that we have solved more of the main theoretical
issues.

\bibliographystyle{plain}
\bibliography{reference}

\end{document}